\documentclass[12pt,preprint]{aastex}

\begin{document}

\title{Chemical Compositions of Kinematically Selected Outer Halo Stars
\altaffilmark{\star}}

\author{Lan Zhang\altaffilmark{1,2}, Miho Ishigaki\altaffilmark{3}, Wako Aoki\altaffilmark{4,5}, Gang Zhao\altaffilmark{1}, and
Masashi Chiba\altaffilmark{3}}

\altaffiltext{$\star$}{Based on data collected at the Subaru
Telescope, which is operated by the National Astronomical
Observatory of Japan}
\altaffiltext{1}{National Astronomical
Observatories, CAS, 20A Datun Road, Chaoyang District, 100012,
Beijing, China; gzhao@bao.ac.cn, zhanglan@bao.ac.cn}
\altaffiltext{2}{Graduate University of the Chinese Academy of
Sciences, 19A Yuquan Road, Shijingshan District, 100049, Beijing,
China}
\altaffiltext{3}{Astronomical Institute, Tohoku University, Sendai, 980-8578, Japan;
   miho@astr.tohoku.ac.jp, chiba@astr.tohoku.ac.jp}
\altaffiltext{4}{National Astronomical
Observatory of Japan, 2-21-1 Osawa, Mitaka, Tokyo 181-8588, Japan;
aoki.wako@nao.ac.jp}
\altaffiltext{5}{Department of Astronomical
Science, School of Physical Sciences, The Graduate University of
Advanced Studies (SOKENDAI), 2-21-1 Osawa, Mitaka, Tokyo 181-8588,
Japan}

\begin{abstract}

Chemical abundances of 26 metal-poor dwarfs and giants are
determined from high-resolution and high signal-to-noise ratio
spectra obtained with Subaru/HDS. The sample is selected so that
most of the objects have outer-halo kinematics. Self-consistent
atmospheric parameters were determined by an iterative procedure
based on spectroscopic analysis. Abundances of 13 elements,
including $\alpha$-elements (Mg, Si, Ca, Ti), odd-Z light elements
(Na, Sc), iron-peak elements (Cr, Mn, Fe, Ni, Zn) and
neutron-capture elements (Y, Ba), are determined by two independent
data reduction and LTE analysis procedures, confirming the
consistency of the stellar parameters and abundances results. We
find a decreasing trend of [$\alpha$/Fe] with increasing [Fe/H] for
the range of $-3.5 <$ [Fe/H]$ < -1$, as found by Stephens and
Boesgaard (2002). [Zn/Fe] values of most objects in our sample are
slightly lower than the bulk of halo stars previously studied. These
results are discussed as possible chemical properties of the outer
halo in the Galaxy.

\end{abstract}

\keywords{Galaxy: abundances - Galaxy:halo - Galaxy: evolution -
stars: abundances}

\section{Introduction}

In recent years, comprehensive studies of stars in the Galactic old
components, such as the halo and the thick disk, have been carried
out to understand the origin of our Galaxy and its early evolution.
A mainstream approach is to get such information through studying
the detailed elemental abundances of metal deficient stars, because
many of these objects have formed from primordial gas clouds during
the early chemo-dynamical evolution of the Galaxy (e.g., Beers \&
Christlieb 2005 and references therein). Indeed, in the course of
Galactic chemical evolution, heavy elements synthesized during the
massive star evolution and the supernova explosion are mixed into
interstellar matter, from which subsequent generations of stars are
formed (e.g., Shigeyama \& Tsujimoto 1998). Extremely metal-poor
stars that we are currently observing are candidate low-mass stars
among those formed in this early stage of the Galaxy. Chemical
abundance studies for such metal-poor stars in the past few decades
(e.g., Zhao \& Magain 1990, 1991; McWilliam et al. 1995; Gratton et
al. 1997; McWilliam 1998; Carretta et al. 2002; Cayrel et al. 2004;
Barklem et al. 2005) and modeling of massive stars and supernova
explosions (e.g., Heger \& Woosley 2002; Umeda \& Nomoto 2005)
provide useful constraints on the nucleosynthesis processes in the
early Galaxy.

These old stellar populations often show high space motions relative
to the stars in the thin-disk component, which reflects the early
dynamical motion of the Galaxy, such as an overall collapse and/or
merging process. Thus, studies on kinematics of metal-poor stars
provide unique information on Galactic structure and formation. For
instance, \cite{chi00} studied the space motions of about 1200
metal-poor stars with [Fe/H]$<-0.6$, utilizing the accurate
measurements of their proper motions by the Hipparcos mission. Their
results have revealed characteristic kinematics of stars, such as a
discontinuous change of mean rotational motion ($<V_{\phi}>$) at
[Fe/H]$\sim -1.7$\footnote{ [A/B] = $\log(N_{\rm{A}}/N_{\rm{B}}) -
\log(N_{\rm{A}}/N_{\rm{B}})_{\odot}$}, suggesting a discontinuous
evolution of the Galaxy's collapse between the formation stage of
the halo and disk components, and some kinematical substructures as
indicative of merging debris. Indeed, more evidence of merging of
smaller galaxies in the formation of the Galaxy has been found
\citep{iba94,pre94,bul05}. It has also been made clear that the halo
is divided into two separate components, the flattened inner halo
and spherical outer halo \textbf{(e.g., Sommer-Larsen \& Zhen 1990; Norris
1994)}. More recently, Carollo et al. (2007) showed, based on the
large sample provided by Sloan Digital Sky Survey (SDSS), that the
outer halo exhibits a net retrograde motion in contrast to a small
prograde rotation of the inner halo, indicating that the Galaxy
collapse was not monolithic unless otherwise its spin motion would
have been unidirectional.

Further constraints on the scenario of Galaxy formation can be
obtained from the detailed chemical abundances of individual stars
in the nearby dwarf galaxies, which are possible remnants of
building blocks of the Galactic halo. A key is the abundance ratio
between the $\alpha$ elements and the iron-peak elements. For
instance, while the bulk of halo stars have over-abundances of
$\alpha$ elements, a significant fraction of stars in Galactic dwarf
galaxies  have comparatively low [$\alpha$/Fe] ratios (e.g.,
Shetrone et al. 2001, 2003; Aoki et al. 2009b). However, the halo
component itself holds a dual nature as mentioned above, so further
studies of the relationship between chemical abundance such as
$\alpha$-elements and kinematical properties of individual halo
stars will provide new insight into the formation of the Galaxy.

Analysis of chemical abundances and kinematics of a large sample of
galactic halo stars have been carried out by e.g.,
\cite{ful00,ful02} and \cite{ste02} (hereafter SB02), who discussed
a possible correlation between [$\alpha$/Fe] and kinematical
properties in particular for the halo stars having extreme orbital
motions. While weak correlations between the abundances with some
orbital parameters were suggested, their samples of field halo
stars, including stars having extreme motions, show distinct
chemical abundances from those seen in the nearby dwarf galaxies.
From these results, it was concluded that the accretion of dwarf
galaxies similar to those currently orbiting our Galaxy did not play
a key role in the formation of the Galaxy. However, neither of them
covered adequate number of stars that have large $Z_{\rm max}$, the
maximum distance of the orbit above and below the Galactic plane,
which is one useful indicator to select outer halo stars, thereby
allowing us to investigate any systematic trends in abundances as a
function of orbital parameters. Moreover, no star with [Fe/H] $<
-$2.0 was considered for the analysis of the abundance-kinematics
correlation in the study of Fulbright (2002). Further studies of
candidate outer halo stars covering a wider metallicity range are
needed.

The aim of this work is thus to obtain detailed chemical abundances
of metal-poor halo stars selected based on their orbital motions, so
that the correlation between the abundance ratios and kinematics
properties (i.e., if the star belongs to the inner or outer halo) is
explored. We determine the abundances of elements including Na,
$\alpha$-elements (Mg, Si, Ca, Ti), Sc, iron peak elements (Cr, Mn,
Ni, Zn) and heavy elements (Y, Ba) for 26 metal-poor halo stars.
This work is based on high resolution and high signal-to-noise (S/N)
ratio spectra from the Subaru Telescope as described in Section 2.
The determination of atmospheric parameters and the procedures of
abundance analysis are described in Section 3 and Section 4,
respectively. In addition, the results for the derived elemental
abundances and related discussion are given in Section 5 and Section
6. The summary is given in the last section.

\section{Observations and data reduction}

\subsection{Selection of stars}

In order to study the chemical abundances of outer halo stars, we
selected candidate metal-poor stars for which large values of
$Z_{\rm max}$ have been obtained from catalogues of \cite{bee00},
\cite{car94} and \cite{rya91}. Orbital parameters, $R_{\rm apo}$
(the maximum radial distances from the Galactic center) and $Z_{\rm
max}$, were calculated from radial velocities, proper motions and
distance in the method described in \cite{chi00}. Although SB02 have
studied candidates of outer halo stars, their sample includes only a
few stars having large $Z_{\rm max}$. In order to study the high
$Z_{\rm max}$ range, we include red giant stars in our sample. Our
sample also includes three stars that were studied by SB02 for
comparison purposes.

Based on this selection, high resolution spectra were obtained for
32 stars (see \S~2.2 for details). However, the S/N ratios are
insufficient for six stars among them for the present purpose.
Hence, the sample of the present work contains 26 stars, among which
16 stars are red giants and the others are main-sequence stars.

Among our sample, BD+04\degr 2466 is reported as a binary member
\citep{jor05}. Carney et al. (2003) reported that BD+01\degr 3070
might be a binary because of its radial velocity variation.
G~112--43 also possibly belongs to a binary system \citep{lat02}.
There is no evidence of binarity for the other stars.

\subsection{High resolution spectroscopy and data reduction}

We obtained high-resolution spectra of our targets using the Subaru
Telescope High Dispersion Spectrograph (HDS; Noguchi et al. 2002) in
February 2003. Table 1 lists the objects and details of the
observations. The spectra cover the wavelength range from 4100 to
6800 {\AA} with a resolving power of 50,000. The CCD on-chip binning
was not applied. Therefore, the resolution element is sampled by
about six CCD pixels.

The standard MIDAS routines were used for data reduction, including
order identification, wavelength calibration, flat-field correction,
background subtraction, and 1D spectra extraction. The Doppler shift
was corrected by measurements for at least 34 moderately strong and
unblended lines before continuum rectification. The S/N ratios (per
0.9~km~s$^{-1}$ pixel) of the reduced spectra at 5800 {\AA} are
presented in Table 1. Examples of a portion of spectra are shown in
Fig. 1.

\subsection{Equivalent widths}

Three different methods were used to measure the equivalent widths:
fitting of a Gaussian profile, fitting of a Voigt profile and a
direct integration. The Gaussian fitting was adopted for weak lines
with apparently no or little blending. The Voigt profile fitting is
applied to strong lines. The direct integration is applied to
moderately strong lines which can be distinguished clearly from
others. The results are presented in Table 2 (electronic version)

The equivalent widths measured by the present work are compared with
the results of SB02 for the three common stars (Fig. 2). Their
spectra were taken with Keck/HIRES, and the spectral quality is
similar to ours (mean S/N $\sim$ 200, R = 30,000 $\sim$ 60,000). We
found that the agreement between the two measurements is fairly
good. The correlations of the two measurements are presented as
follows (dashed lines in Fig. 2):

E.W.s(SB02)$_{\rm G~15-13} = -0.02(\pm0.31)+1.029(\pm0.008) \times$
E.W.s(TW) (m{\AA})

E.W.s(SB02)$_{\rm G~166-37} = 0.14(\pm0.77)+1.002(\pm0.017) \times$
E.W.s(TW) (m{\AA})

E.W.s(SB02)$_{\rm G~238-30} = -0.11(\pm0.44)+0.979(\pm0.022) \times$
E.W.s(TW) (m{\AA})

The standard deviations of these three relations are 3.0 m{\AA}, 4.0
m{\AA} and 1.6 m{\AA}, respectively.

The uncertainty of equivalent width measurement is estimated by
using the formula of \cite{cay88}:
\begin{equation}
\sigma_{w} = \rm \frac{1.5\sqrt{FWHM\delta_{x}}}{S/N}
\end{equation}
where FWHM (in m{\AA}) is the full width at the half maximum of a
line; $\delta_{x}$ is the pixel scale ( m{\AA}/pixel), and S/N is
the signal-to-noise ratio estimated for the spectral order which
contains the absorption lines. For instance, the lowest and highest
S/N ratios are 80 and 520 among these spectra, which lead to the
errors of 0.6 m{\AA} and 0.07 m{\AA}, respectively. Taking the
uncertainty of the continuum rectification into consideration,
$3\sigma_{w}$ is adopted as the detection limit of absorbtion lines.
We estimated that the errors of equivalent width measurements are
0.2 $\sim$ 1.8 m{\AA}, depending on the S/N ratio and the strength
of lines.

\section{Stellar atmospheric parameters}

\subsection{Determination of effective temperature}

The effective temperature ($T_{\mathrm{eff}}$) is determined by
making the abundances of Fe be independent of the excitation
potential of \ion{Fe}{1} lines used in the analysis (see \S~4 for
details of the abundance analysis). An initial value is set in the
code to calculate the iron abundances from neutral species, and then
a slope between $\log \epsilon$(\ion{Fe}{1}) and the excitation
potential is determined by least-square fit. This process is
iterated until the slope is minimized (Fig. 3). \textbf{The
perturbation of $T_{\mathrm{eff}}$ is added to change the slope,
making the slope of the change within 1$\sigma$. Finally, the
perturbations are taken as the uncertainties of $T_{\mathrm{eff}}$,
and they are around 100~K.}

We prefer the spectroscopic method to determine the effective
temperature in the present work, because this can be applied to all
objects in the sample, and many clean \ion{Fe}{1} lines are
available in our high S/N spectra.

The estimate of $T_{\mathrm{eff}}$ by profile fitting for Balmer
lines is not adopted. This technique is sensitive to the continuum
determination for the echelle order containing broad absorption
features of Balmer lines. Moreover, for giant stars, the Balmer line
profiles are not very sensitive to the effective temperature.

The estimate of $T_{\mathrm{eff}}$ from color indices (e.g. $V-K$)
is not adopted to derive the final results in the present work,
because photometry data for our sample are incomplete, and the
$T_{\mathrm{eff}}$ estimate is sensitive to the interstellar
reddening. The error of $T_{\mathrm{eff}}$ determined from $V-K$ due
to the uncertainty of reddening is estimated to be as large as
100~K. Although such uncertainty exists in the method, we estimate
$T_{\mathrm{eff}}$ from color indices for stars for which photometry
data are available for comparison purposes. The $T_{\mathrm{eff}}$
scale for dwarfs from \cite{alo96} and for giants from \cite{alo99}
are adopted, and the colors of the stars were taken from the 2MASS
catalogue for near-infrared photometry and the Hipparcos catalogue
\citep{per97} for optical one. The dust maps of \cite{sch98} are
employed to estimate the reddening values. The $E(\bv)$ values are
corrected taking into account the finite distance to a star, by
using the method described in \cite{bee00}.

Since the Johnson system and the Telescope Carlo S\'{a}nchez system
(TCS) were adopted in the work of \cite{alo96,alo99}, the JHK
indices from 2MASS are transformed through the relations of
\cite{ram04}:

\begin{equation}
J_{\rm{TCS}} = J_{\rm{2MASS}}+0.001-0.049(J-K)_{\rm{2MASS}}
\end{equation}
\begin{equation}
H_{\rm{TCS}} = H_{\rm{2MASS}}-0.018+0.003(J-K)_{\rm{2MASS}}
\end{equation}
\begin{equation}
K_{\rm{TCS}} = K_{\rm{2MASS}}-0.014+0.034(J-K)_{\rm{2MASS}}
\end{equation}

Comparisons of $T_{\mathrm{eff}}$ derived from the color indices
(hereafter CI, the average value derived from
$T_{\mathrm{eff}}^{V-K}, T_{\mathrm{eff}}^{J-H}$ and
$T_{\mathrm{eff}}^{J-K}$) with those from the spectroscopic analysis
(hereafter SA) are listed in Table 3, and are shown in Fig. 4. The
average and standard deviation of the difference is:
$<T_{\mathrm{eff}}^{\rm{CI}}-T_{\mathrm{eff}}^{\rm{SA}}> = 54$ and
$\sigma = 99$. The difference is also given by linear least-square
fit: $T_{\mathrm{eff}}^{\rm{CI}} = 20 + 1.01 \times
T_{\mathrm{eff}}^{\rm{SA}}$

These comparisons indicate that the effective temperatures estimated
from color indices for our sample are in fairly good agreement with
those determined by the spectroscopic method in general.

\subsection{Determination of other parameters}

\emph{Surface gravity}. $\log g$ is determined by forcing
\ion{Fe}{1} and \ion{Fe}{2} to give the same iron abundance.
In addition, Ti ionization equilibrium is also used to check the
final results. Although this method is affected by non-LTE effects
and uncertainties of atomic data, no other method provides reliable
gravity for the whole sample. The uncertainties of Hipparcos
parallaxes are larger than 40\% for a half of our sample stars. The
gravity is not well determined from Y2 isochrone for giants
\citep{coh02}. Therefore, ionization equilibrium of iron is used in
the present analysis.  \textbf{In order to estimate the uncertainty
of the $\log g$ determination, we calculate the $\log g$ value that
results in 0.1~dex discrepancy in Fe abundances from Fe I and Fe II.
The typical error in log g derived by this calculation is 0.32~dex.
We note that the typical Fe abundance error (random error) is
0.05~dex ($\S$4.2), so the above estimate of the log g error is
rather conservative.}

 \emph{Metallicity}. The solar
abundance of $\log \varepsilon$(Fe)$_{\odot}$ = 7.51 \citep{and89}
is adopted to get the [Fe/H] and [X/Fe] values.

\emph{Micro-turbulent velocity}. $\xi_{t}$  is also determined from
the abundance analysis. The abundances are derived from individual
\ion{Fe}{1} lines whose equivalent widths are in the range of 10
$\sim$ 100 m{\AA} by changing the $\xi_{t}$ value until these
abundances are independent of their equivalent widths (Fig. 3).
\textbf{With same method described in $\S$3.1, the error of
$\xi_{t}$ is estimated. Typically, the error is 0.3~dex.} The
determination of stellar parameters including $T_{\mathrm{eff}}$ is
iterated until a consistent parameter set is obtained. The
parameters of our sample determined by the above method are listed
in Table 4. As an example, the Fe abundances derived from individual
\ion{Fe}{1} lines are plotted as functions of excitation potential,
equivalent width, and wavelength for HD 108577 in Fig. 3.

\section{Elemental abundance analysis}

The grid of the flux constant, homogeneous, LTE model atmospheres by
Kurucz (1993) in which convection of overshoot approximation
\citep{cas97} is considered is used in our abundance analysis.
Abundance is calculated with the program ABONTEST8, developed by Dr.
Pierre Magain at Li\`{e}ge, Belgium, by requiring the calculated
equivalent width to agree with the observed value. Natural
broadening, thermal broadening, van der Waals damping, and
macro-turbulent velocity are all included in the calculation. The
mean abundance of each element is derived from all available lines
by giving equal weight to each line. The solar compositions of
\cite{and89} were adopted to derive relative abundances.

\subsection{Atomic data}

The atomic line data for Na, Mg, Si, Ca, Ti, Cr, Fe, Ni, Y and Ba
are adopted from \cite{ste99} and SB02. The references for
individual lines are given in their papers. The $\log gf$ values of
Sc, Mn and Zn are taken from \cite{law89}, \cite{boo84} and
\cite{bie80}, respectively.

Absorption lines of Sc, Mn and Ba are known to be influenced by the
hyperfine splitting (HFS). The splitting effect of these elements on
abundance determination is largest for strong lines in which
individual components are partially saturated. The HFS effect is
smaller in weak lines and fully saturated lines. Our analysis
includes the HFS effects using the line data of \ion{Sc}{2} and
\ion{Mn}{1} from \cite{mcw95}, and references therein. The HFS data
of \ion{Ba}{2} are taken from \cite{mcw98} assuming the solar system
isotopic mix given by \cite{sne96}.

\subsection{Estimation of uncertainties}

We estimate the uncertainties in the abundance determination for the
two sources. One is the uncertainties in the analysis of individual
lines, including random errors of equivalent widths, oscillator
strengths, and damping constants; the other is the error due to the
uncertainties of atmospheric parameters.

\subsubsection{Errors from internal uncertainties}

The typical uncertainty in the equivalent width measurement is about
0.6 m{\AA} as mentioned in \S2.3. This results in an error of about
0.04 dex in the elemental abundance calculation from an unblended,
moderately strong line. The scatter of the abundance results from
individual lines gives another estimate of the uncertainty from
equivalent widths and other factors. The error is estimated by
dividing the standard deviation of the derived abundances from
individual lines by a square root of the number of lines used
($N^{\frac{1}{2}}$). The error is negligible when the number of
lines used in the analysis is large. For instance, the standard
deviation ($\sigma$) of the abundance results from 95 \ion{Fe}{1}
lines of G~112--43 is 0.06~dex, which is comparable with the
estimate from equivalent width errors, and the random error ($\sigma
\cdot N^{-\frac{1}{2}}$) is 0.006~dex. For elements for which only a
small number of lines are available (e.g. Mg, Si), the $\sigma$ of
\ion{Fe}{1} (typically 0.05~dex) is adopted in stead of the $\sigma$
of those species.

\subsubsection{Errors from the uncertainties of atmospheric parameters}

From the discussion of $\S 3.1$, the error of $T_{\mathrm{eff}}$ is
estimated to be on the order of 100 K. Considering uncertainties in
Fe line analysis, the uncertainties of $\log g$ and micro-turbulent
velocity are estimated to be 0.3~dex and 0.3~km$\cdot$s$^{-1}$,
respectively. \textbf{These values are equivalent to the average
error of the parameters.} The abundance changes by the changes of
atmospheric parameters are listed in Table 5 for two stars in our
sample: G~166--37 (main sequence) and HD~ 175305 (giant). We note
that the abundance ratios of most elements ([X/Fe] values) are
relatively insensitive to variations of stellar parameters.

\textbf{Finally, the abundance uncertainties is estimated by summing
the atmospheric and internal uncertainties in quadrature.}

\subsection{Comparisons with an independent analysis}

For our dataset, an independent abundance analysis is made by one of
the authors (M.I.) using a different model atmospheres and an
abundance analysis code, in order to demonstrate the reliability of
the final results. Comparisons of the atmospheric parameters and
chemical abundances between the two analyses are presented in Table
6, Table 7 and Fig. 5.

For the independent analysis (SA II), the data reduction is made
using the standard IRAF routines. Equivalent widths are measured
fitting Gaussian profiles to clean absorption lines. The agreement
of the equivalent widths between the two measurements is excellent;
the root mean square (RMS) of the differences is $2.3\pm0.1$ m{\AA}
on average. Stellar atmospheric parameters and abundance analysis
were independently performed with an LTE abundance analysis code
described in \cite{aok09} using the model atmosphere grid of Kurucz
(1993) calculated with NEWODF assuming no convective overshooting
(Castelli \& Kurucz 2003). The HFS effects are taken into account in
the estimate of Ba abundances. Effective temperatures, surface
gravities, micro-turbulent velocities, and [Fe/H] are calculated
with iterative process, as described in \S~3. The resulting values
of the atmospheric parameters reasonably agree each other:

$T_{\mathrm{eff}}^{\rm{SA II}} = 53 + 0.99 \times
T_{\mathrm{eff}}^{\rm{SA}}$

$\log g^{\rm{SA II}} = -0.19 + 1.02 \times \log g^{\rm{SA}}$

[Fe/H]$^{\rm{SA II}} = 0.01 + 1.02 \times$ [Fe/H]$^{\rm{SA}}$

$\xi^{\rm{SA II}} = -0.02 + 1.07 \times \xi^{\rm{SA}}$

The chemical abundance ratios determined by the two independent
analysis agree within $\sim$0.1 dex. We note that the [Fe/H] values
derived by the second analysis are systematically lower by about
0.05~dex. This small difference can be explained by the difference
of the model atmosphere grid. The Kurucz's models assuming no
overshooting are systematically cooler than those assuming
overshooting, resulting in lower Fe abundances.

\subsection{Comparison with previous work}

The three common stars G~15--13, G~166--37 and G~238--30 in SB02,
whose parameters were derived by the same analysis method as ours,
are used for comparisons. Comparisons of atmospheric parameters are
given in Table 8.

Comparisons of the abundance results derived by our analysis with
those of SB02 are presented in Table 9 (the first three lines) for
the three stars in common. The agreement between the two studies is
fairly good in general. The [Fe/H] of G~238--30 derived in the
present work is 0.16~dex higher than that of SB02. This is partially
because we adopt 100~K higher $T_{\mathrm{eff}}$ in the analysis. It
can result in a 0.08~dex difference. G~238--30 is most
metal-deficient in our sample, and the number of iron lines used in
the parameter determination is the smallest. It causes a error of
0.02~dex.\textbf{ Besides, because of the relatively low S/N ratio
and the weak iron lines of this star, the uncertainty of equivalent
width is $\sim$ 2 m{\AA}, which can lead to the error of $\sim$
0.05~dex.} These could be the main reasons for the comparatively
large discrepancy in $T_{\mathrm{eff}}$ and [Fe/H] between the two
studies. The [Na/Fe] of G~15--13 and G~166--37 derived by our
analysis is higher than those of SB02. The primary reason for this
discrepancy might be the difference of equivalent widths. Our
calculation shows that the derived Na abundance increases by
0.07~dex for the increase of equivalent widths by 1~m{\AA}. For
G~15--13, our equivalent widths of \ion{Na}{1} $\lambda\lambda$5682
and 5688 are 0.5 $\sim$ 2.6 m{\AA} higher than those in SB02, while,
for G~116--37, the value of \ion{Na}{1} 5682 is 2.4 m{\AA} larger
and that of 5688 is 0.7 m{\AA} smaller than those in SB02. Such
differences at least partially explain the differences of the Na
abundances between the two studies for the two stars. The effects of
equivalent width errors are significant, because of the weakness of
the Na lines, compared with the lines of other elements. \textbf{The
situation is similar for [Ti/Fe] of G~238--30.}

In order to confirm the consistency of the abundance analysis
technique with model atmospheres, we determined stellar parameters
and abundances using the equivalent widths presented in the paper of
SB02 for three stars having intermediate metallicity (G~5--19,
G~9--36 and G~215--47). The comparisons of the results are given in
Table 9 (the middle three lines). Our re-analysis well reproduces
the results of SB02, confirming the consistency between the two
analyses. A small systematic difference in [Mg/Fe] between the two
works is found, that is, our results are systematically lower by
0.04 $\sim$ 0.11 dex than those in SB02.

Finally, the analysis for the same stellar parameters adopting the
equivalent widths of SB02 are attempted, and the results are
compared with those of SB02 in Table 9 (the bottom three lines). The
agreement is excellent, confirming the consistency of the abundance
analysis code between the two studies.

\section{Chemical abundance results}

The derived abundances by the present analysis are summarized in
Table 10 - 14. We show the abundance ratios of elements with respect
to Fe as a function of metallicity in Fig. 6.

\subsection{$\alpha$-elements}

In our analysis, abundances of $\alpha$-elements are measured from 3
lines of magnesium, 3 lines of silicon, 23 lines of calcium, and 33
lines of titanium for most stars.

The average value and standard deviation of abundance ratios for
each element are as follows: [Mg/Fe] = +0.31$\pm$0.15 dex (25
stars), [Si/Fe] = +0.36$\pm$0.17 dex (20 stars), [Ca/Fe] =
+0.29$\pm$0.15 dex (26 stars) and [Ti/Fe] = +0.27$\pm$0.13 dex (26
stars). The average of these abundance ratios for the four elements
($<$[$\alpha$/Fe]$>$) derived from our sample is +0.30$\pm$0.08 for
giants and +0.29$\pm$0.13 for dwarfs. The behaviors of magnesium,
silicon and calcium are similar: an increasing trend of their
relative abundance with decreasing metallicity can be seen in Fig.
6, while the average abundance ratio of Ti is lower than those of
other three elements (see bottom panel of Fig. 7). A possible
interpretation is that Ti is formed during complete and incomplete
silicon burning while Si and Ca are produced by incomplete explosive
silicon and oxygen burning \citep{cay04}.

In Fig. 7, we show our results for the four $\alpha$-elements along
with previous abundance studies (SB02; Gratton et al. 2003). Most of
the previous studies for metal-poor stars that are mainly belonging
to the inner halo concluded that [$\alpha$/Fe] is constant for
[Fe/H] $< -1$. \cite{mcw95} and \cite{rya96} showed that
[$\alpha$/Fe] $\sim$ +0.4 dex for [Fe/H] $< -2$, while \cite{gra97}
gave a mean [$\alpha$/Fe] (+0.26$\pm0.08$) for most halo stars
through self-consistent analysis of 100 metal-poor ([Fe/H] $< -0.5$)
dwarfs. On the other hand, our results of [$\alpha$/Fe] values show
a slope for [Fe/H] $< -1$, as found by SB02. The slope of
[$\alpha$/Fe] versus [Fe/H] found in our analysis is $-0.14$, which
is similar to the value found by SB02 ($-$0.15) for 55 metal-poor
halo stars. The dependence of chemical abundance trend on the
kinematics properties is discussed in \S~6.

\subsection{Light odd-Z elements}

\subsubsection{Sodium}

For stars with metallicities of [Fe/H] $> -2.5$, four lines at
$\lambda$$\lambda$5682, 5688, 6154 and 6160 were used for sodium
abundance determination. The deviations from LTE formation for these
lines are less than 0.1 dex \citep{bau98}, which is smaller than the
typical errors in our analysis (0.1~dex). Hence, the non-LTE effects
on these lines can be neglected. However, for some extremely
metal-poor stars in our sample, these lines are too weak to be
detected. Therefore, resonance lines at $\lambda$$\lambda$5890 and
5896 were included to compute the sodium abundances for stars with
[Fe/H] $< -2.5$. Non-LTE effects of these two lines in metal-poor
dwarfs cannot be ignored \citep{bau98}. The abundance values derived
by resonance lines are corrected by the estimate of non-LTE effects
by \cite{geh04,geh06}.

\cite{geh06} analyzed sodium abundances with non-LTE correction for
55 metal-poor dwarf stars. In the upper panel of Fig. 8, our results
are compared with those in their study. For dwarfs the two sets of
results agree well, while the $\rm [Na/Fe]_{NLTE}$ values of giants
in our sample are $\sim$ 0.35~dex larger than in dwarfs in $-2.7
\lesssim$ [Fe/H] $\lesssim -1.7$. In several giants, the initial
sodium abundance may be changed because of the products of the Ne-Na
cycle from deeper layers being dredged to the surfaces
\citep{and07}; The [Na/Fe] in some giants might be explained by this
effect, although a small systematic errors in the [Na/Fe] ratios for
giants and dwarfs are not excluded.

\subsubsection{Scandium}

In Fig. 6, the average of [Sc/Fe] ratios of dwarfs is greater than
the solar abundance ratio, while for giants there is a slowly
decreasing trend with decreasing metallicity with a scatter of
0.14~dex. In the bottom panel of Fig. 8, our results are compared
with those in \cite{zhao90}. They studied 20 metal-poor dwarfs, and
found the overabundance of scandium relative to iron. A weak
decreasing trend with decreasing metallicity for [Sc/Fe] of giants
is noticed.  The [Sc/Fe] of dwarfs are statistically 0.15~dex larger
than that of giants in the similar metallicity range, mostly because
of the high [Sc/Fe] values derived by \cite{zhao90}. Similar
conclusion was also reported very recently by \cite{bon09}.

\subsection{Iron peak elements}

Abundances of four iron-peak elements, Cr, Mn, Ni and Zn, were
determined in our present paper.

The results of [Cr/Fe] derived from neutral lines in our analysis
show a decreasing trend with decreasing metallicity for giants. This
confirms the results of previous studies for low-metallicity stars
\citep{mcw95, gra03, cay04, bar05}. Our results are shown with the
previous work in the upper panel of Fig. 9. In Fig. 10, Cr
abundances from \ion{Cr}{2} lines and logarithic abundance
difference of Cr from \ion{Cr}{2} and \ion{Cr}{1} lines ([Cr~$_{\rm
II}$/Cr~$_{\rm I}$]) are plotted as functions of [Fe/H]. The
decreasing abundance ratio of Cr with decreasing [Fe/H] is not found
in [Cr/Fe]$_{\rm{II}}$. The average [Cr~$_{\rm II}$/Cr~$_{\rm I}$]
is 0.32~dex for giants and 0.28~dex for dwarfs. Similar offset
between neutral and ionize Cr was also reported in \cite{lai08}.
\cite{sob07} suggested that this discrepancy may be due to NLTE
effects. Besides, the decreasing trend found for Cr in giants is not
seen in dwarfs studied by SB02 and \cite{gra03}. Such a result is
also found by \cite{lai08} and \cite{bon09}. Possible explanations
for this discrepancy were discussed by \cite{lai08}, and is not
repeated here.

Our [Mn/Fe] presents a plateau at --0.44~dex in $-2.5 <$ [Fe/H] $<
-1$. In the second panel of Fig. 9, combining with the halo dwarf
samples of \cite{gra03}, a $\sim$ 0.16~dex systematic difference of
[Mn/Fe] between dwarfs and giants in the metallicity range of $-2 <$
[Fe/H] $< -1$ is noticed. Although we lack of enough data for lower
metallicity range ([Fe/H] $< -3$), \cite{bon09} presented that such
phenomena are also found in extreme metal-poor stars. They also
suggested that the discrepancy of these two elements between giants
and dwarfs may be reduced if 3D models are adopted.

The Ni abundance ratios shown in Fig. 6 are approximately constant
around the solar value. $<$[Ni/Fe]$>$ = $-$0.06$\pm$0.08 dex. This
result agrees well with the study of \cite{ste99} ($<$[Ni/Fe]$>$ =
$-$0.09$\pm$0.07) and SB02 ($<$[Ni/Fe]$>$ = $-$0.06$\pm$0.20).
\cite{tsu95} mentioned that [Ni/Fe] yields in the mass-averaged SNe
II are roughly consistent with solar abundance ratios. 

Several previous works conclude that [Zn/Fe] $\sim$ 0 for -2.5
$\lesssim$ [Fe/H] $<$ 0, while at a lower metallicity, the ratio
increases with declining of [Fe/H] \citep{cay04,bar05,ume05}. In the
comparison with \cite{bar05} in the bottom panel of Fig. 9, we found
that these features are not clear in our sample because of the
existence of low [Zn/Fe] stars in [Fe/H] $< -2.5$. Moreover, the
[Zn/Fe] of our sample is lower than other stars studied by previous
work in general. This suggests the low Zn abundance in outer halo
stars. This possibility is discussed in $\S$6 in more detail.

\subsection{Neutron capture elements}

The last two panels of Fig. 6 show the abundance ratios [Y/Fe] and
[Ba/Fe] as a function of [Fe/H]. Comparisons with previous works are
presented in Fig. 11. Ba is also chosen as the reference element to
investigate the [Y/Ba] (Fig. 12).

[Y/Fe] and [Ba/Fe] of BD+04\degr 2466 are obviously higher than
other stars. We estimated the carbon abundance of this star as
[C/Fe] = 1.21 dex. As mentioned in $\S$2.3, this star is believed to
belong to a binary system because of its radial velocity variation.
Thus, this star would have obtained s-process-rich material from an
AGB companion \citep{bus01}.

Both [Y/Fe] and [Ba/Fe] increase with increasing [Fe/H] in the lower
metallicity range ([Fe/H] $< -2$), while the ratios are
approximately constant ([Y/Ba] $\sim$ 0) at the higher metallicity.
At lower metallicity, [Y/Ba] increases with decreasing [Fe/H] or
[Ba/H]. The behaviors of Y and Ba are similar for [Fe/H] $> -2$. By
contrast, no object having high [Y/Fe] or [Ba/Fe] is found in our
sample in [Fe/H] $< -$2. Such under-abundance of neutron-capture
elements in the low metallicity range is possibly a chemical
property of outer halo stars. However, given the small number of
stars studied here, we cannot derive any conclusion.

The bottom panel of Fig. 12 shows the [Y/Ba] as a function of
[Ba/H]. This can be divided into two parts: near solar abundance for
[Ba/H] $> -2.5$ and increasing of [Y/Ba] with barium decreasing at
lower [Ba/H]. \cite{hon04} and \cite{aok05} discussed that the
``weak r-process'' which contributes only to the light
neutron-capture elements affects most stars with [Fe/H] $\gtrsim
-3.5$. The feature in Fig. 12 indicates that the stars with [Ba/H]
$< -2.5$ ($-3.25 <$ [Fe/H] $< -2.25$) in our sample have been
significantly affected by this process.

\subsection{Magnesium as the metallicity indicator}

Although the Fe abundance is adopted as the metallicity indicator in
the above discussion as usual, the Mg abundance may provide better
estimate of metallicity, because that is produced by hydrostatic
burning processes in massive stars, while Fe production is sensitive
to the supernova processes yet understood well. \cite{shi98} and
\cite{cay04} also recommended that choosing this element as
metallicity tracer. Thus, the abundances of odd-Z elements and
$\alpha$-elements relative to magnesium vs. [Mg/H] are plotted in
Fig. 13. It is clear that the behavior of Ti is different from other
$\alpha$-elements (Fig. 13). The reason is discussed in $\S$5.1.

In the first plot of Fig. 13, we notice that a relatively large
scatter in [Na/Mg] is seen in giants which are in the metallicity
range of [Mg/H] $< -1.5$, while the correlation of [Na/Mg] vs.
[Mg/H] is tighter in the higher metallicity range ([Mg/H] $> -$1.5).
As discussed in $\S$5.2.1, some giants with large sodium abundances
enhanced by dredging process cause higher [Na/Mg] and, consequently,
result in large scatter of [Na/Mg] at lower [Mg/H].

A strong metallicity dependence of abundance ratios of the odd-Z
elements (Na, Sc) is seen in Fig. 13. The production of Na is
thought to be controlled by the neutron flux which depends on the
metallicity of the SN. Therefore, a statistic increase of [Na/Mg]
with increasing [Mg/H] is expected. Besides, Sc is a product of
explosive oxygen and neon burning \citep{woo95}, the enhanced
mechanism of the Sc is thought to be the surplus of neutrons in
$^{22}$Ne which is formed from $^{14}$N by the CNO circle
\citep{kob06}. The low abundance ratios of Sc in low [Mg/H] is the
result of small amount of CNO elements.

\section{Discussion}

As mentioned in \S 5.1, the [$\alpha$/Fe] vs. [Fe/H] diagram
obtained for our sample stars presents a finite slope, namely,
decreasing [$\alpha$/Fe] with increasing [Fe/H] (Fig. 6). A similar
trend was also reported by SB02, who studied candidate outer-halo
stars. Fig. 7 shows the abundance ratios of $\alpha$-elements for
our sample and that of SB02 as well as of \cite{gra03}. The stars of
the present work and of SB02 show the decreasing trend and the
presence of a large scatter in the [$\alpha$/Fe] ratios at
[Fe/H]$>-2$, compared with the objects studied by \cite{gra03}, most
of which are inner halo stars.
Thus, although the [$\alpha$/Fe] ratios (in particular [Mg/Fe]) of
the halo sample, usually dominated by the inner halo population,
have been conventionally regarded to be constant, this is not the
case for the outer halo stars.

Our measurements also suggest that the [Zn/Fe] ratios of our sample
are systematically lower than other halo stars previously studied.
This element was not studied by SB02. Most of our objects with
$-2<$[Fe/H]$<-1$ have sub-solar [Zn/Fe] values, while the bulk of
halo stars previously studied (most of them would belong to the
inner halo) have [Zn/Fe]$\sim 0$. The three most metal-poor stars
([Fe/H]$<-2.5$) in our sample have [Zn/Fe]$\sim 0$, lower than the
trend found for other stars in this metallicity range previously.

We here investigate a possible correlation between abundance ratios
and orbital parameters of our current sample stars. Figures 14 and
15 show the abundance ratios for Mg, Si, Ca and Zn as functions of
$Z_{\rm{max}}$ and $R_{\rm{apo}}$, respectively. In these plots,
small and large marks correspond to [Fe/H]$<-2$ and $>-2$,
respectively. As is evident, no clear trend of [X/Fe] against the
orbital parameters is found for the three $\alpha$ elements. For
[Fe/H]$<-2$ the [$\alpha$/Fe] ratios remain larger than those for
[Fe/H]$>-2$ and a star with highest [$\alpha$/Fe] is located at
large $Z_{\rm{max}}$ or $R_{\rm{apo}}$. This suggests that
metal-poor, outer halo stars with [Fe/H]$<-2$ are largely enriched
by SNe II, without any dependence on their orbital parameters. Also
the outer halo stars with [Fe/H]$>-2$ show roughly constant
[$\alpha$/Fe] ratios with increasing $Z_{\rm{max}}$ or
$R_{\rm{apo}}$. We note that HD~134439 having the lowest
[$\alpha$/Fe] has low $Z_{\rm{max}}$, but has very high
$R_{\rm{apo}}$, and is classified into the outer halo. In contrast
to [$\alpha$/Fe], one may find a weak decreasing trend of [Zn/Fe]
with increasing $Z_{\rm{max}}$ and $R_{\rm{apo}}$. However, the
sample size is yet too small to derive any definitive conclusion for
this trend.

These results indicate that, among our sample stars which reside in
the outer halo component, there is no significant dependence of
chemical abundances on kinematical parameters {\it within} this
population. There is a signature of some systematic, chemical
difference between the inner and outer halo: smaller [$\alpha$/Fe]
with large scatters at [Fe/H]$>-2$ (Fig. 7) and smaller [Zn/Fe] at
[Fe/H]$<-2$ (Fig. 9) for the outer halo than the inner one. However,
in order to set tighter constraints on these chemical trends, a much
larger sample of both inner and outer halo stars is clearly
required.

The decreasing trend of [$\alpha$/Fe] with increasing metallicity
found for the outer halo stars recalls the chemical nature of dwarf
spheroidal galaxies (dSphs) around the Galaxy (e.g., Shetrone et al.
2001, 2003).  \cite{lan04} studied abundance ratios of six dSph
galaxies, and predicted a plateau for lower metallicity range
([Fe/H] $\lesssim -1.8$) and a sudden decrease for [Fe/H] $\gtrsim
-1.8$ in the relationship between [$\alpha$/Fe] and [Fe/H]. The
production of $\alpha$ elements are mainly from SNe II explosions in
a short time scale while Fe can be from SNe Ia and SNe II. The sharp
decline of [$\alpha$/Fe] implies intense galactic winds and
contributions of SNe Ia to Fe from [Fe/H]$\sim -1.8$ in dSph
galaxies \citep{lan04}. If a merging and collision of certain dSph
galaxies happened before significant contributions of SNe Ia, it
should result in normal [$\alpha$/Fe] in outer halo of the Galaxy.
Indeed, in the simulation of \cite{joh08}, they suggested that
$\alpha$-rich stars in outer halo may come from dSph galaxies by
merging at early epoch. On the other hand, mergers of dSph galaxies
at later times would result in low [$\alpha$/Fe] stars in the outer
halo.

The lower [Zn/Fe] values were also reported for stars belonging to
the nearby dSph studied by Shetrone et al. (2001, 2003). Among their
sample, majority of dSphs stars having metallicity of
$-2<$[Fe/H]$<-1$ show [Zn/Fe]$<0$ similar to our sample of outer
halo stars.  For the more metal-poor stars ([Fe/H]$<-2$) in our
sample, [Zn/Fe] approximately follow the solar value as discussed in
Section 5.3, except for one star showing [Zn/Fe] $> 0.4$.  This
behavior is similar to metal-poor stars in the ultra-faint dwarf
spheroidal galaxies studied in Frebel et al. (2009). However, the
sample size of stars in dwarf galaxies for which the Zn abundance is
studied is too small to derive any conclusion. Further measurements
of Zn abundances for dwarf galaxy stars as well as for outer halo
objects would be a key to understanding the contributing of massive
progenitors to the metal-enrichment in these systems.

We also remark that the [Mn/Fe] ratios of the outer halos stars
having an approximately constant value of $\sim -$0.4~dex agree well
with those obtained for dSph stars (e.g., Shetrone et al. 2001,
2003). Thus, combined with the properties of [$\alpha$/Fe] and
[Zn/Fe] ratios discussed above, it is consistent with the hypothesis
that the outer halo is largely made up with late merging/accretion
of dSphs which are similar to those currently observed.

\textbf{In the analysis of \cite{roe09}, a large spread (0.5 --
0.7~~dex) in [Ni/Fe] for the outer halo stars was noticed. However,
in our sample, the scatter of [Ni/Fe] of outer halo stars is
0.05~dex, which is similar to that of our inner halo ones
(0.06~dex). The abundance ratios of \cite{roe09} were collected from
different sources, therefore, the large scatter in his analysis is
partially caused by different spectral qualities, different methods
of stellar parameter determination and different structure of model
atmospheres. Our consistency results do not support the claim about
the difference of the abundance scatter between the inner and outer
halo populations. Thus, a sample with larger size and from single
source is needed to probe whether Ni can be taken as another
indicator for chemical inhomogeneity between inner and outer halo
populations.}

\section{Summary}

Elemental abundances and kinematics of 26 metal-poor halo stars in
$-3.5 <$ [Fe/H] $< -1.0$ were studied with high-resolution and high
signal-to-noise ratio spectra taken from Subaru/HDS. Most objects
have large values of $Z_{\rm max}$ and/or $R_{\rm apo}$, indicating
their outer halo population.

1. The $\alpha$-elements (Mg, Si, Ca, Ti) to iron ratios are
overabundant relative to the solar values, and increase with
decreasing [Fe/H]. From the investigation of [$\alpha$/Fe] with
kinematics, it is concluded that the slope of [$\alpha$/Fe] vs.
[Fe/H] is caused by outer halo stars. This result bears a
resemblance to the abundance trend of $\alpha$-elements found in
dSph galaxies around the Milky Way, suggesting contributions of
mergers of dSph galaxies at later times to the formation of the
outer halo structure.

2. There exist low [Zn/Fe] stars in the very low metallicity range,
which is different from the trend of Zn abundances found by previous
studies, but is rather similar to those of ultra-faint dwarf
galaxies. A weak slope is displayed in the plots of [Zn/Fe] against
kinematic parameters. These signatures should be confirmed by future
work based on a larger sample.

3. The neutron-capture elements Y and Ba are under-abundant,
following the trend found for the bulk of field halo stars. No
neutron-capture enhanced object was found in our sample, and that is
a possible property of outer halo stars, though studies for a larger
sample is also needed.  Increasing Y with decreasing Ba in lower
[Ba/H] indicates that the stars with lower [Fe/H] and [Ba/H] in our
sample have experienced the ``weak r-process''.

4. Discrepancies of elemental abundances by the order of 0.15~dex
between giant and dwarf stars are found for Sc, Cr and Mn. Our
results support the recent study for these elements by \cite{bon09},
though the reasons for these discrepancies are still unclear. A
discrepancy of 0.35~dex between giants and dwarfs is found for
[Na/Fe] in [Fe/H]$<-2$ even after non-LTE corrections, possibly
reflecting abundance variations in several giants from their
original composition.

\acknowledgments

LZ thanks Drs. J. R. Shi, Y. Q. Chen and J. S. Deng for useful
suggestions and discussions. This work is supported by the NSFC
under grant 10821061 and by the National Basic Research Program of
China under grant 2007CB815103. MC acknowledges support from a
Grant-in-Aid for Scientific Research (20340039) of the Ministry of
Education, Culture, Sports, Science and Technology in Japan.

\clearpage
\begin{figure}
\includegraphics[angle=-90,scale=.70]{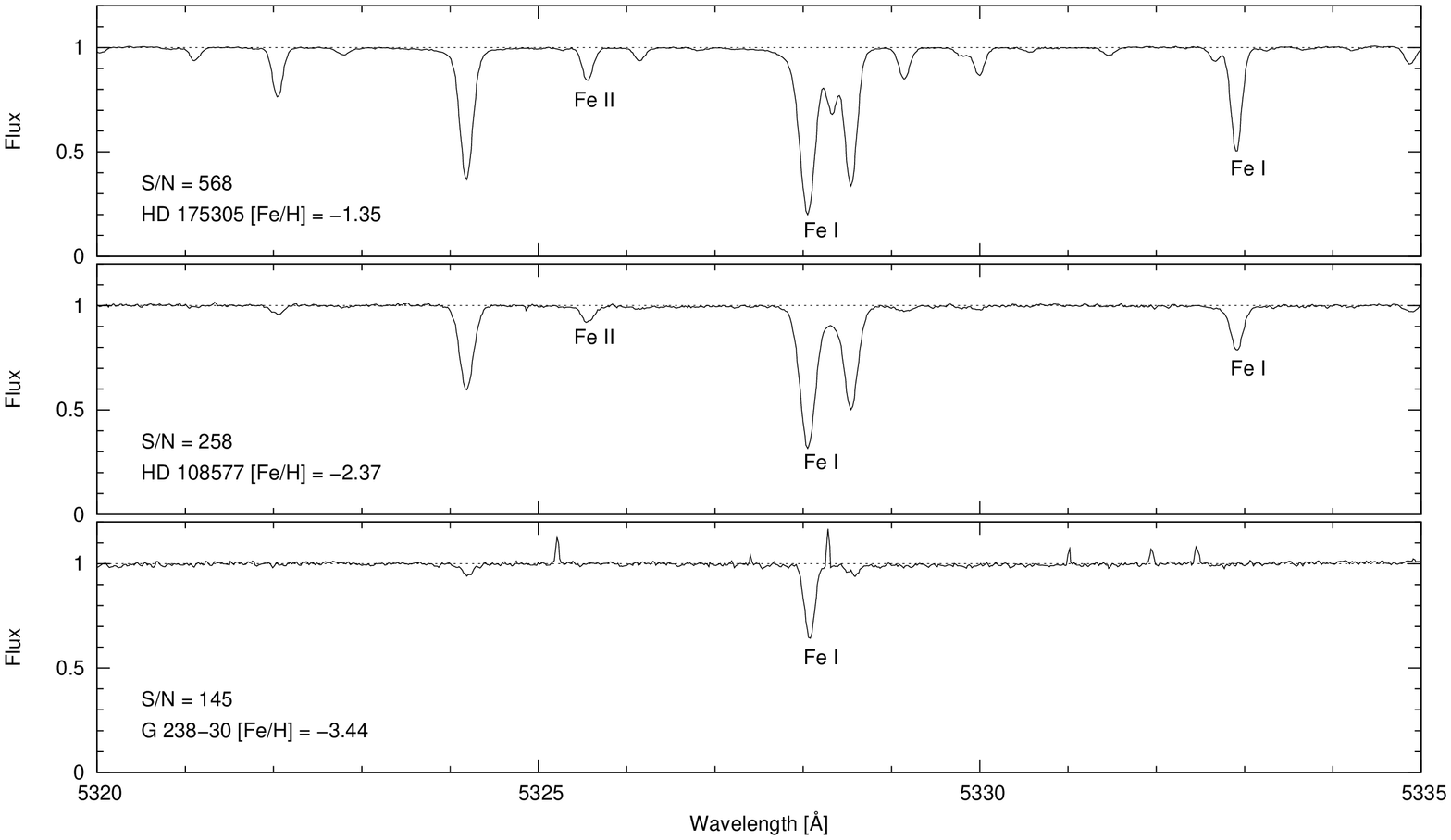}
\figcaption{Examples of spectra obtained with Subaru/HDS for HD
175305 ($T_{\mathrm{eff}}$ = 5035, $\log g$ = 2.84, [Fe/H] =
$-$1.35) with S/N $\sim$ 568, HD 108577 ($T_{\mathrm{eff}}$ = 4800,
$\log g$ = 1.13, [Fe/H] = $-$2.37) with S/N $\sim$ 270 and G~238-30
($T_{\mathrm{eff}}$ = 5490, $\log g$ = 3.57, [Fe/H] = $-$3.44) with
S/N $\sim$ 130. }
\end{figure}

\begin{figure}
\epsscale{0.4} \plotone{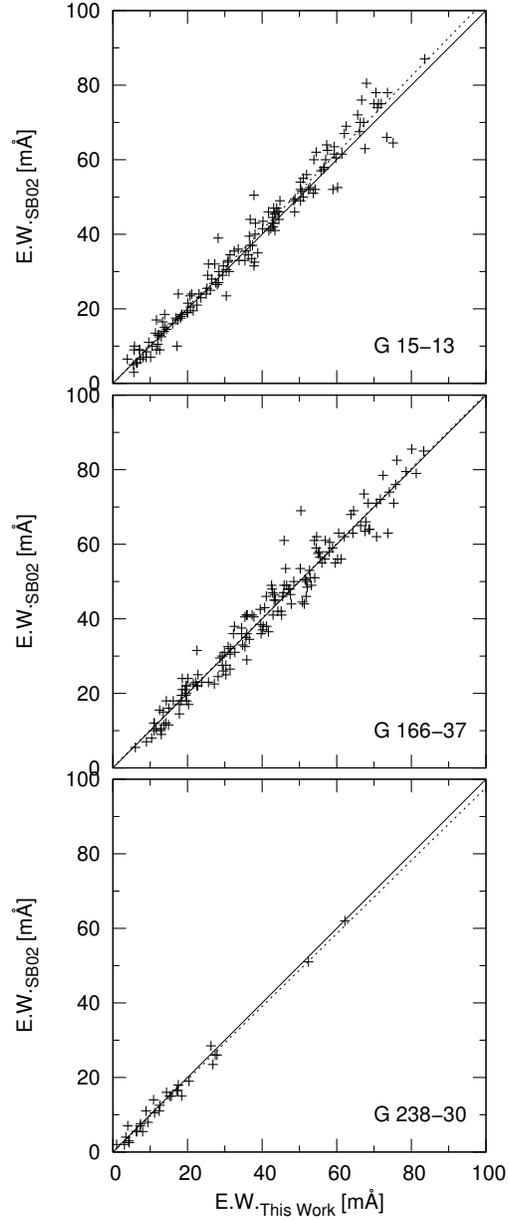} \figcaption{Comparison of the
equivalent widths measured in this work with those of SB02. In all
cases the solid line
  represents a one-to-one correlation and the dotted line represents a linear fit of the data.}
\end{figure}

\begin{figure}
\epsscale{0.5} \plotone{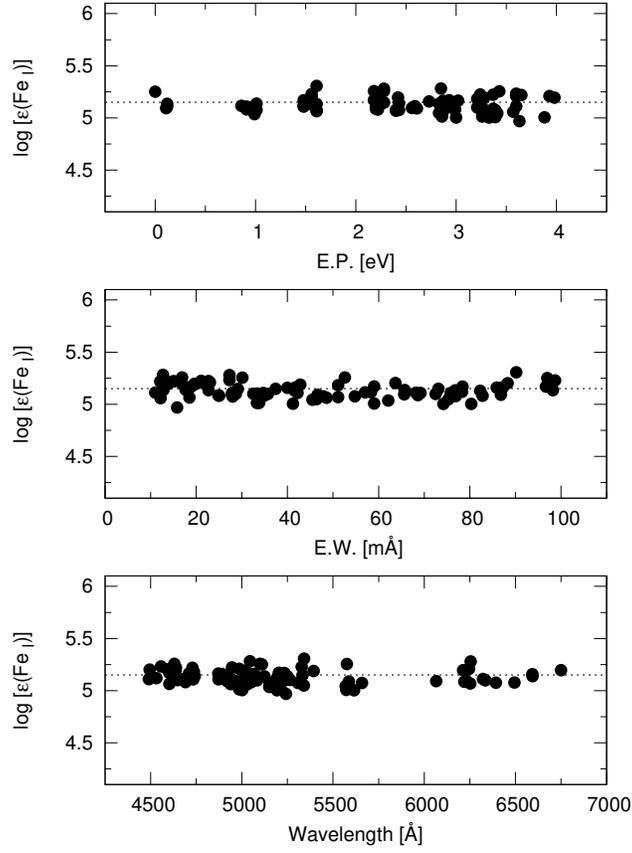} \figcaption{\ion{Fe}{1} abundances
derived from individual \ion{Fe}{1} lines for HD 108577 as functions
of excitation potential (E.P.), equivalent width (E.W.), and
wavelength. The dashed line represents the mean value of iron
abundance. The parameters are $T_{\mathrm{eff}}$ = 4800K, $\log g$ =
1.13, [Fe/H] = $-$2.37, $\xi_{t}$= 1.8 km$\cdot$s$^{-1}$. }
\end{figure}

\begin{figure}
\centering
\includegraphics[angle=-90,scale=0.7]{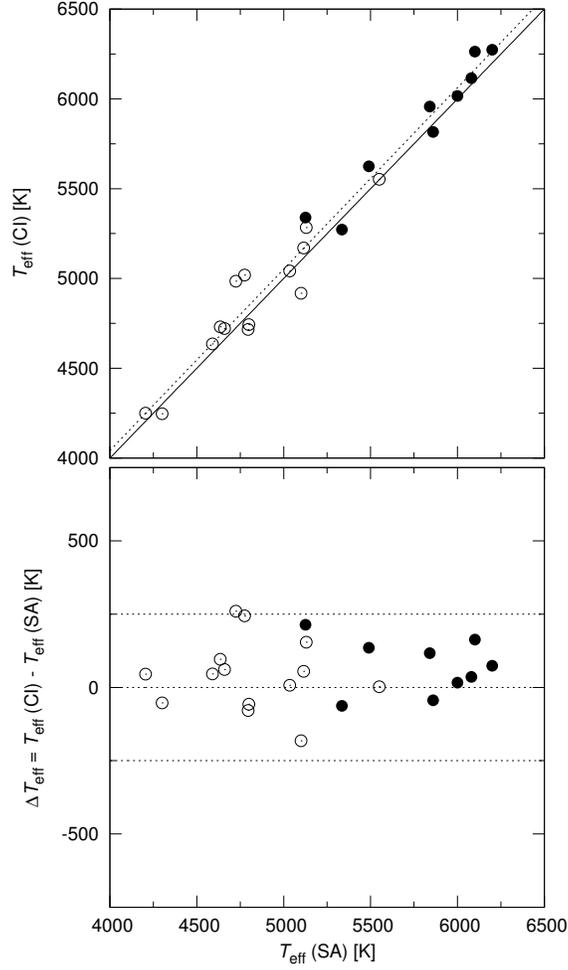}\figcaption{Comparison of
$T_{\mathrm{eff}}$ adopted finally (SA) vs. those derived by color
indices (CI). The solid line represents a one-to-one correlation
while the dashed line represents a linear fit of the data. Filled
circles represent dwarfs while open ones mean giants.}
\end{figure}

\begin{figure}
\centering
\includegraphics[angle=-90,scale=0.7]{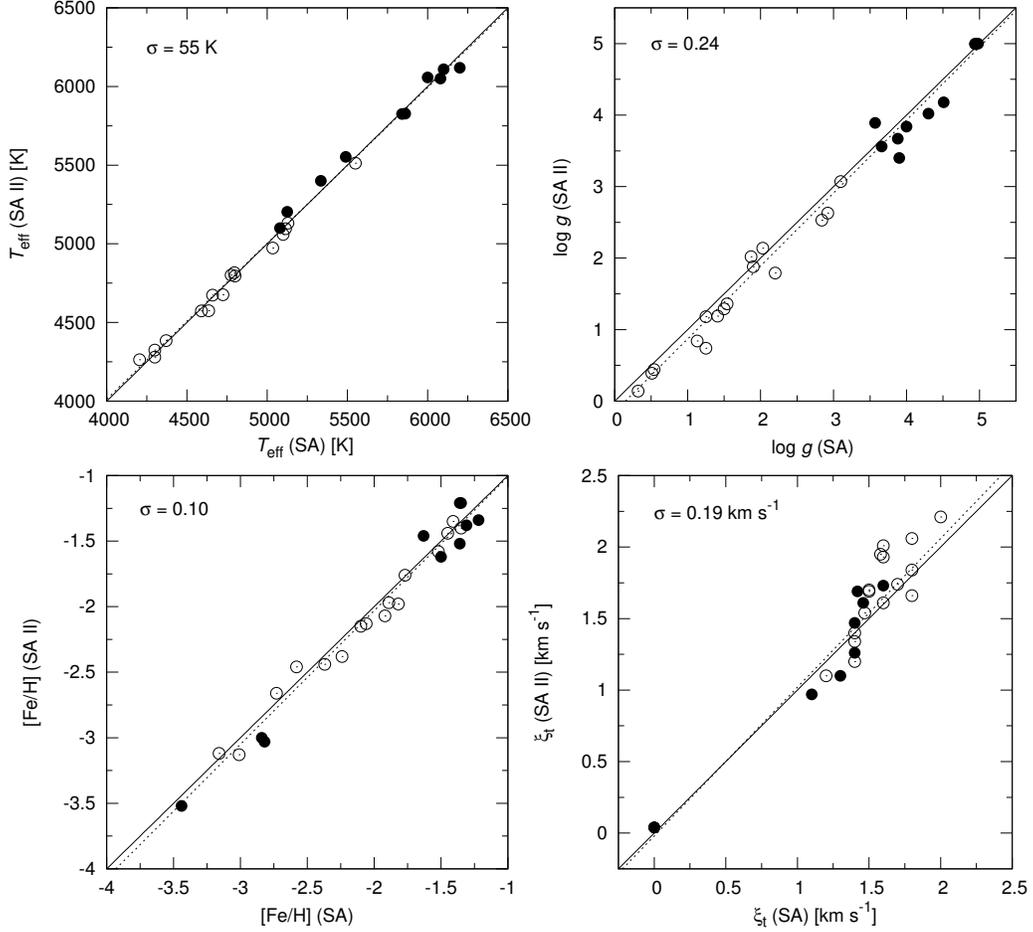}\figcaption{Comparison of
atmospheric parameters adopted finally (SA) with those derived by an
independent analysis (SA II; see text). The solid line represents a
one-to-one correlation while the dashed line represents a linear fit
of the data. The symbols are the same as in Fig. 4.}
\end{figure}

\begin{figure}
\epsscale{1} \plotone{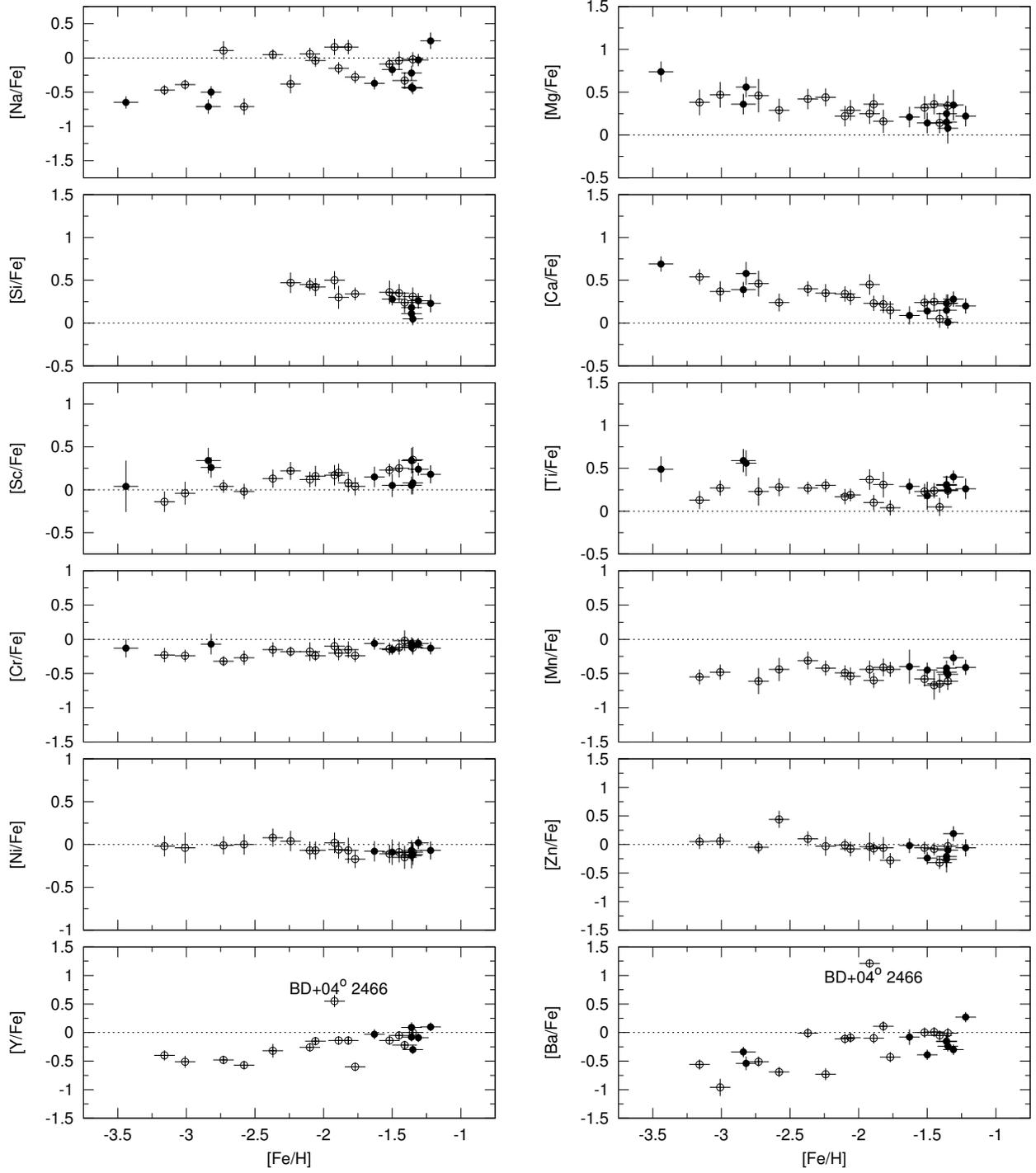} \figcaption{Plots of the elemental
abundance ratios against metallicity. The symbols are the same as in
Fig. 4.}
\end{figure}

\begin{figure}
\epsscale{0.5} \plotone{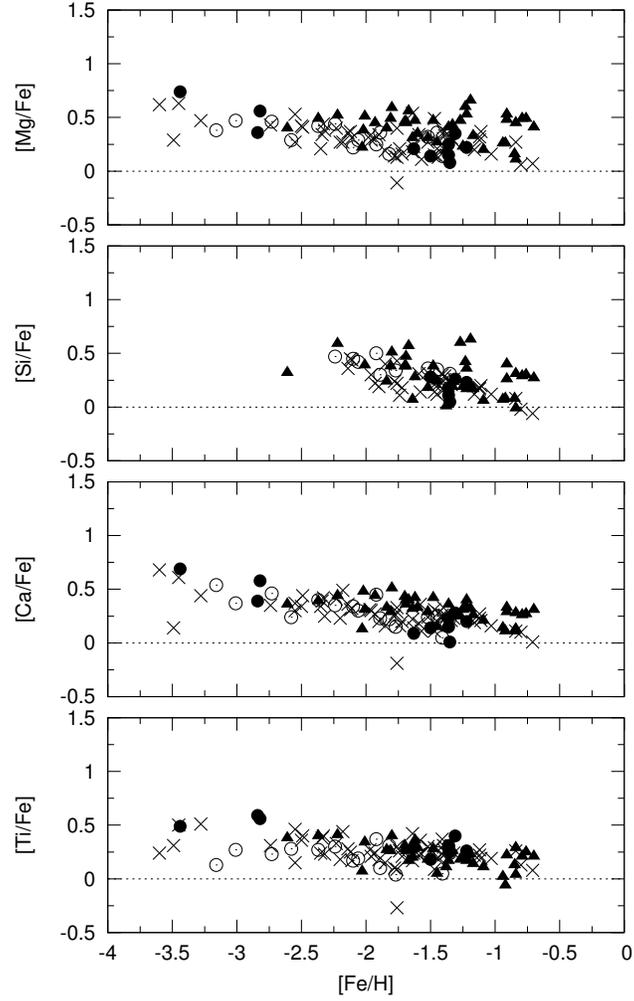} \figcaption{A plot of our observed
[$\alpha$/Fe] values (circles, same as Fig. 4) and previous
abundance studies. The crosses represent the results of SB02, while
filled triangles mean the results of halo samples from
\cite{gra03}.}
\end{figure}

\begin{figure}
\epsscale{0.5} \plotone{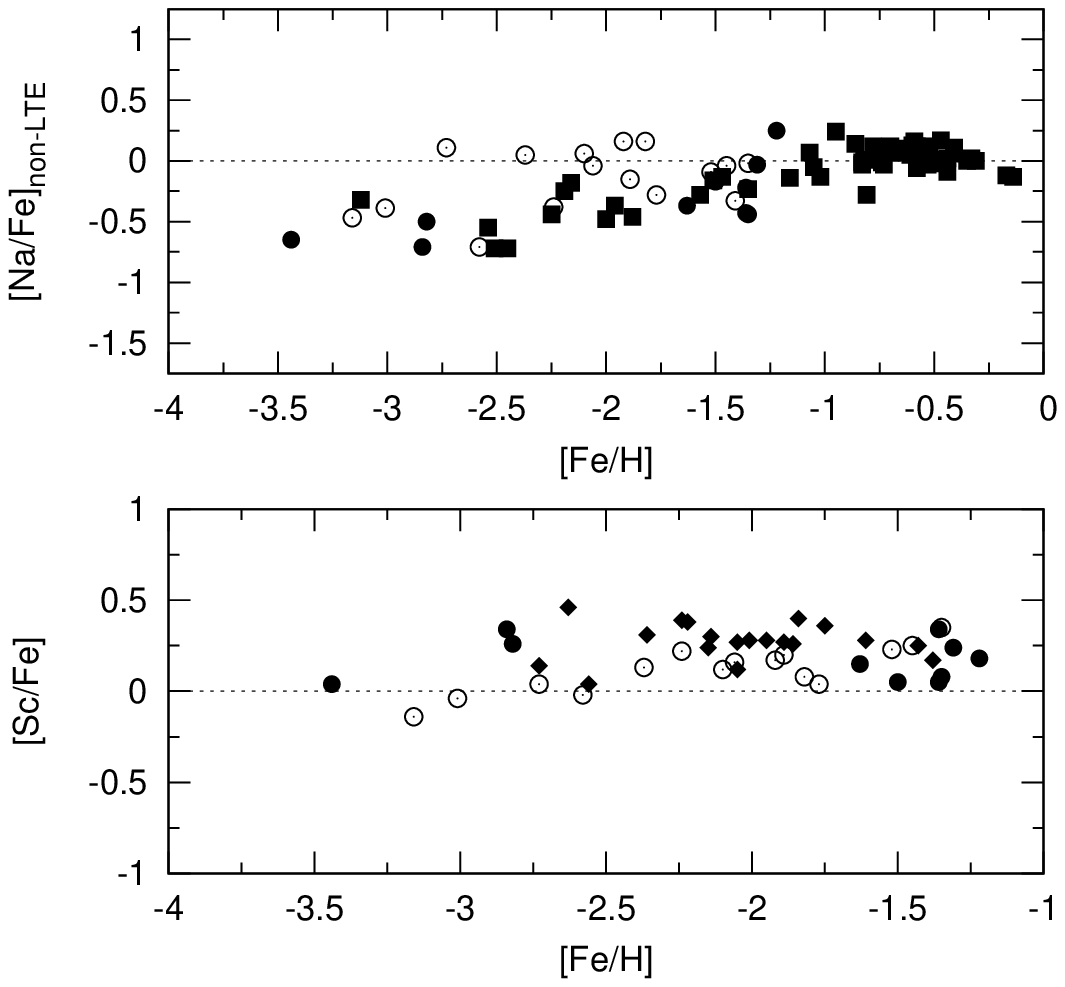}\figcaption{Same as Fig. 7, but for
[Na/Fe] and [Sc/Fe]. Filled squares are dwarfs from \cite{geh06},
and filled diamonds are dwarfs from \cite{zhao90}.}
\end{figure}

\begin{figure}
\epsscale{0.5} \plotone{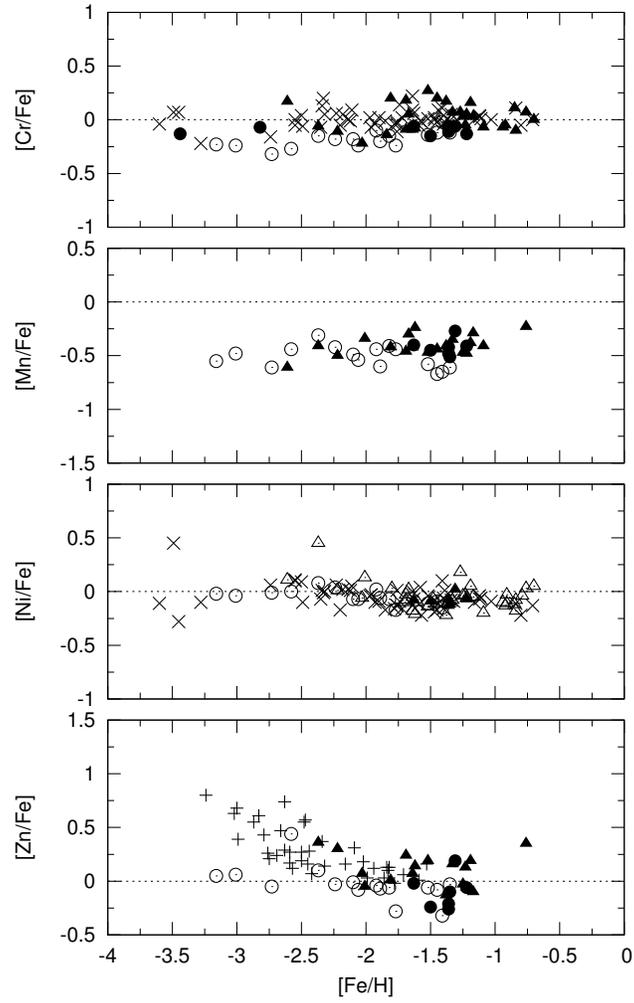} \figcaption{Same as Fig. 7, but for
Fe-peak elements. Pluses represent the results of \cite{bar05}. }
\end{figure}

\begin{figure}
\epsscale{0.5} \plotone{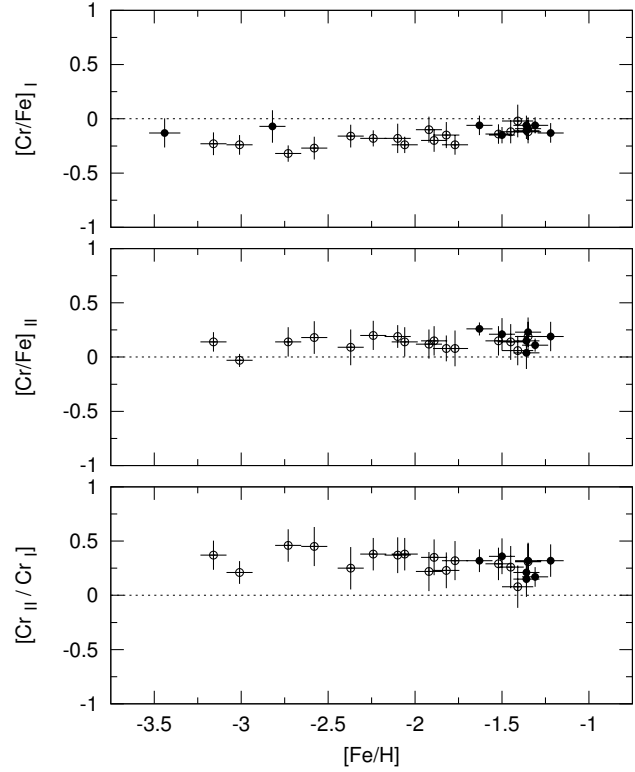} \figcaption{Cr abundance derived
from \ion{Cr}{1} and \ion{Cr}{2} ([Cr/Fe] $_{\rm{II}}$) against with
[Fe/H], along with [\ion{Cr}{2}/\ion{Cr}{1}]. The symbols are the
same as in Fig. 4. }
\end{figure}

\begin{figure}
\epsscale{0.5} \plotone{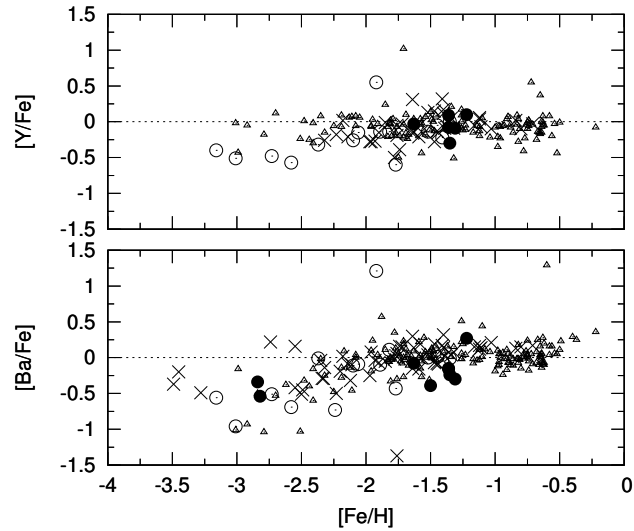} \figcaption{Same as Fig. 7, but for
[Y/Fe] and [Ba/Fe]. Small open triangles represent the results from
\cite{ful00}. }
\end{figure}

\begin{figure}
\epsscale{0.5} \plotone{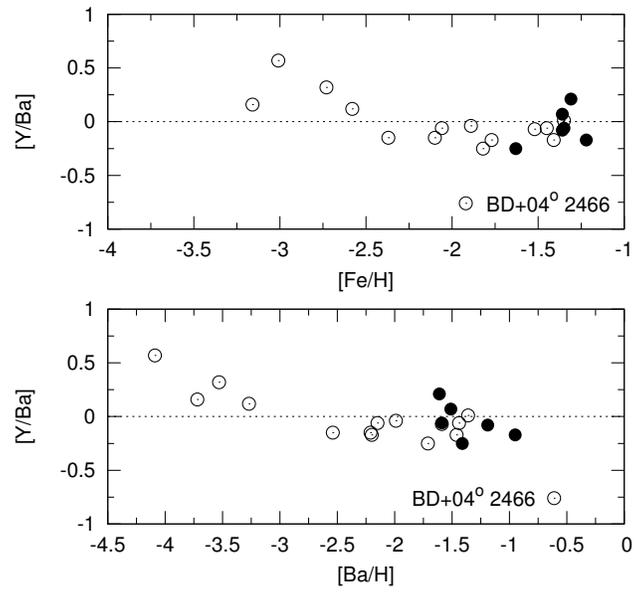} \figcaption{A plot of [Y/Ba] vs.
[Fe/H] and [Ba/H]. The symbols are the same as in Fig. 4.}
\end{figure}

\begin{figure}
\epsscale{0.5} \plotone{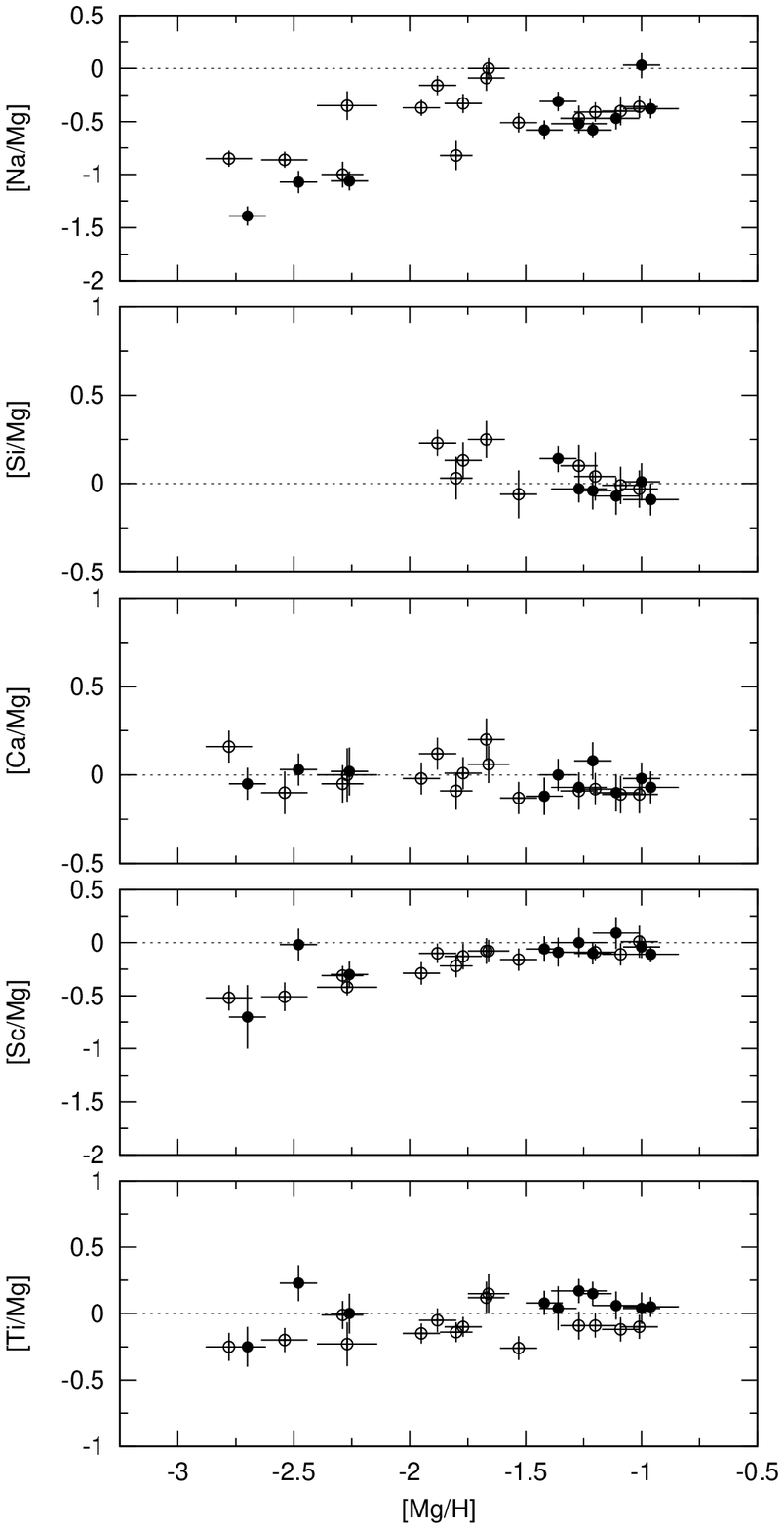} \figcaption{Same as Fig. 6, but for
[(Na, Si, Ca, Sc, Ti)/Mg] vs. [Mg/H]. }
\end{figure}

\begin{figure}
\epsscale{0.5} \plotone{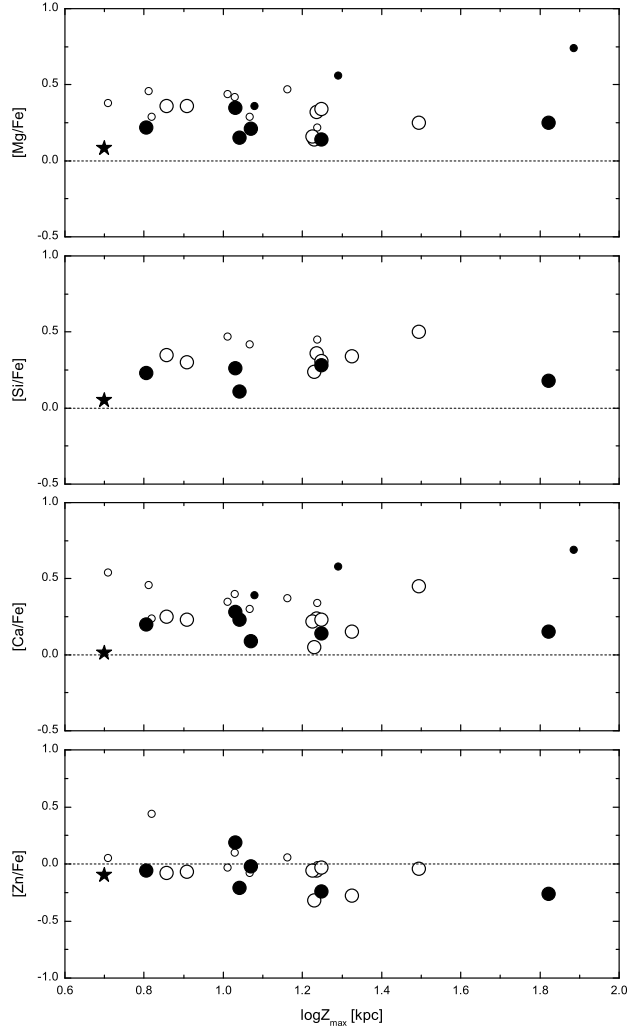} \figcaption{Plots of [$\alpha$/Fe]
and [Zn/Fe] against the logarithm of $Z_{\rm{max}}$. The symbols are
the same as in Fig. 4. The black star is HD~134439 which has low
$Z_{\rm{max}}$, but it is classified into the outer halo because of
its very high $R_{\rm{apo}}$. Besides, the large circle means the
stars with [Fe/H] $>$ $-$2, while small ones are those with [Fe/H]
$<$ $-$2. }
\end{figure}

\begin{figure}
\epsscale{0.5} \plotone{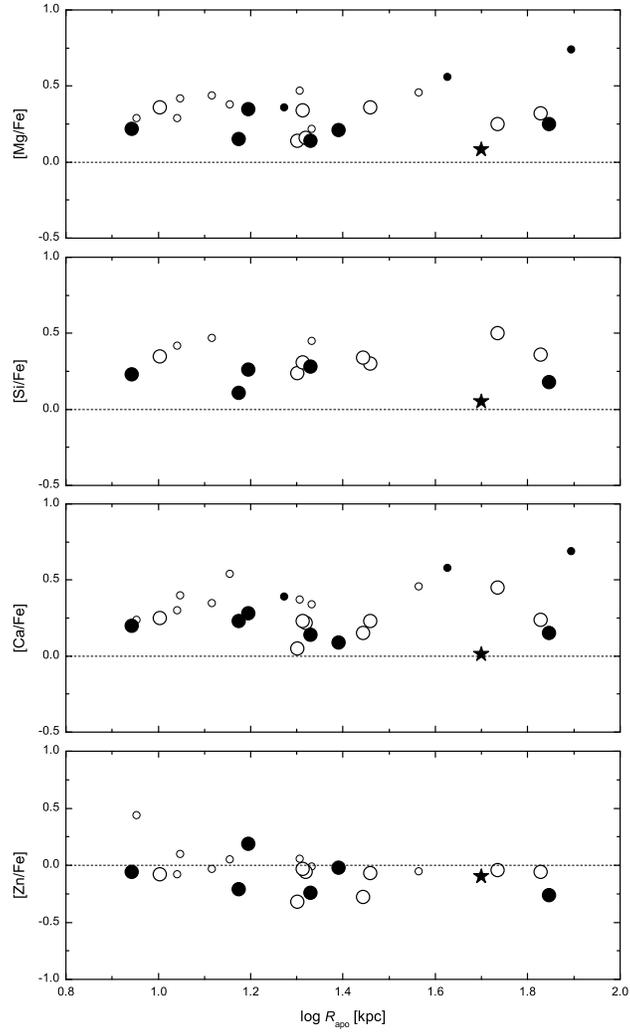} \figcaption{Plots of [$\alpha$/Fe]
and [Zn/Fe] against the logarithm of $R_{\rm{apo}}$. The symbols are
the same as in Fig. 14. }
\end{figure}


\clearpage
\begin{deluxetable}{llccrcrrrrr}
\tabletypesize{\scriptsize}
\tablewidth{0pt}
\tablecaption{Log of the Subaru/HDS } \tablehead{ \colhead{Code} &
\colhead{Name} & \colhead{RA} & \colhead{DEC} &
\colhead{\textit{V}$^{a}$} & \colhead{[Fe/H]$^{a}$} & \colhead{Exp.}
& \colhead{\textit{N}$_{\rm photon}$$^{b}$} &
\colhead{\textit{v}$_{\rm rad}$$^{c}$} &
\colhead{$R_{\rm{apo}}$} & \colhead{$Z_{\rm{max}}$} \\
\colhead{} & \colhead{} & \colhead{(J2000)} & \colhead{(J2000)} &
\colhead{} & \colhead{} & \colhead{[s]} & \colhead{} &
\colhead{[km$\cdot$s$^{-1}$]} & \colhead{[kpc]} & \colhead{[kpc]}}
\startdata
 1 & BD+04\degr~2466   & 11 26 49.20 & $+$03 51 52.2 & 10.53 & $-$1.88 & 1200 &   30511 &      37.1 & 54.26 & 31.20 \\
 2 & BD+01\degr~3070   & 15 22 40.08 & $+$01 15 52.9 & 10.06 & $-$1.85 &  900 &   35921 &  $-$327.4 & 67.32 & 17.19 \\
 3 & BD+09\degr~2870   & 14 16 29.98 & $+$08 27 52.9 &  9.40 & $-$2.39 &  900 &   40356 &  $-$119.7 & 36.62 &  6.48 \\
 4 & BD+10\degr~2495   & 12 59 19.96 & $+$09 14 35.4 &  9.72 & $-$1.83 &  600 &   36539 &     251.5 & 21.49 & 17.30 \\
 5 & BD+12\degr~2547   & 13 04 06.59 & $+$11 26 16.3 &  9.92 & $-$2.07 &  900 &   23928 &       6.6 & 28.79 &  8.09 \\
 6 & BD+29\degr~2356   & 13 01 52.42 & $+$29 11 17.8 & 11.50 & $-$1.06 & 3600 &   20992 &  $-$207.3 & 10.06 &  7.19 \\
 7 & BD+30\degr~2611   & 15 06 53.82 & $+$30 00 37.0 &  9.13 & $-$1.32 &  600 &   64797 &  $-$280.4 & 19.98 & 16.98 \\
 8 & HD~33771          & 05 10 49.56 & $-$37 49 02.9 &  9.45 & $-$1.93 & 1500 &   29398 &   $-$12.8 & 10.97 & 11.67 \\
 9 & HD~85773          & 09 53 39.25 & $-$22 50 08.2 &  9.42 & $-$2.27 & 1200 &   38485 &     148.2 &  8.97 &  6.59 \\
10 & HD~107752         & 12 22 52.75 & $+$11 36 25.8 & 10.01 & $-$2.74 & 1200 &   42834 &     220.4 & 14.29 &  5.11 \\
11 & HD~108577         & 12 28 16.91 & $+$12 20 41.5 &  9.57 & $-$2.56 &  600 &   38485 &  $-$111.0 & 11.15 & 10.67 \\
12 & HD~119516         & 13 43 26.76 & $+$15 34 31.2 &  9.05 & $-$2.49 &  600 &   51786 &  $-$284.5 & 20.86 & 16.78 \\
13 & HD~124358         & 14 13 21.38 & $-$12 09 23.9 &  9.50 & $-$1.98 &  600 &   38284 &     324.7 & 27.77 & 21.12 \\
14 & HD~128279         & 14 36 48.47 & $-$29 06 43.6 &  8.02 & $-$2.20 &  300 &   55752 &   $-$75.1 & 13.04 & 10.24 \\
15 & HD~175305         & 18 47 05.73 & $+$74 43 30.8 &  7.18 & $-$1.45 &  300 &  165037 &  $-$184.0 & 20.56 & 17.67 \\
16 & HD~237846         & 09 52 38.68 & $+$57 54 58.7 &  9.93 & $-$2.63 & 1200 &   23208 &  $-$303.1 & 20.24 & 14.49 \\
17 & HD~134439         & 15 10 13.09 & $-$16 22 45.4 &  9.09 & $-$1.57 &  600 &   63772 &     310.4 & 50.08 &  5.01 \\
18 & G~112--43         & 07 43 43.97 & $-$00 04 00.9 & 10.22 & $-$1.51 &  900 &   27629 &   $-$83.8 & 15.67 & 10.74 \\
19 & G~115--58         & 09 10 48.10 & $+$46 22 36.6 & 12.08 & $-$1.74 & 2400 &   14492 &     226.8 & 14.91 & 11.00 \\
20 & G~15--13          & 15 12 32.43 & $+$06 01 44.3 & 12.32 & $-$1.71 & 5400 &   23527 &     219.7 & 24.59 & 11.76 \\
21 & G~166--37         & 14 34 51.00 & $+$25 09 54.0 & 12.67 & $-$1.47 & 5400 &   17456 &     369.4 & 70.11 & 66.39 \\
22 & G~238--30         & 13 17 40.00 & $+$64 15 00.0 & 12.91 & $-$2.96 & 5400 &   17825 &     229.7 & 78.38 & 76.89 \\
23 & G~41--41          & 09 29 15.56 & $+$08 38 00.6 & 11.15 & $-$2.80 & 1800 &   19291 &     266.9 & 18.75 & 11.97 \\
24 & G~48--29          & 09 40 43.20 & $+$01 00 29.6 & 10.48 & $-$2.66 & 1200 &   19565 &   $-$56.4 & 42.30 & 19.53 \\
25 & G~53--41          & 10 27 24.25 & $+$01 24 00.1 & 11.04 & $-$1.34 & 1800 &   20473 &      88.2 &  8.76 &  6.38 \\
26 & LP~894--3         & 05 57 50.70 & $-$30 53 48.7 & 11.26 & $-$1.83 & 2400 &   18495 &     305.7 & 21.37 & 17.71 \\
\enddata
\tablecomments{a: $V$ band magnitude and metallicity are taken from
\cite{bee00}, \cite{car94} or \cite{rya91}; b: The signal-to-noise
ratio and number of photon per unit pixel evaluated at
$\sim$5800{\AA}; c: Heliocentric radial velocities.}
\end{deluxetable}

\clearpage
\begin{deluxetable}{lccrcrrrrrr}
\tabletypesize{\scriptsize}
\tablewidth{0pt} \tablecaption{Atomic data and measured equivalent
widths \label{tbl-2}} \tablehead{ \colhead{Element}           &
\colhead{$\lambda$ (\AA)} & \colhead{E.P.}          & \colhead{$\log
gf$}  & \colhead{$\Delta\gamma_{6}$}          & \colhead{1}    &
\colhead{2} & \colhead{3}  & \colhead{4} & \colhead{5} &
\colhead{......} } \startdata
\ion{Na}{1} & 5682.650 & 2.10   & $-$0.82 & 2.0   &   6.9 &  12.1 &   3.0 &   7.1 &   6.9 & ......\\
\ion{Na}{1} & 5688.219 & 2.10   & $-$0.37 & 2.0   &  14.6 &  19.9 &   6.2 &  11.2 &  15.3 & ......\\
\ion{Na}{1} & 5889.951 & 0.00   &    0.12 & 2.0   &   ... &   ... & 196.0 &   ... &   ... & ......\\
\ion{Na}{1} & 5895.924 & 0.00   & $-$0.18 & 2.0   &   ... &   ... & 173.7 &   ... &   ... & ......\\
\ion{Na}{1} & 6154.227 & 2.10   & $-$1.66 & 2.0   &   ... &   1.4 &   2.5 &   ... &   ... & ......\\
\ion{Na}{1} & 6160.751 & 2.10   & $-$1.35 & 2.0   &   1.0 &   ... &   ... &   1.1 &   ... & ......\\
\ion{Mg}{1} & 4571.099 & 0.00   & $-$5.59 & 2.5   &  64.3 &  84.2 &   ... &  81.8 & 116.6 & ......\\
\ion{Mg}{1} & 4702.996 & 4.34   & $-$0.55 & 2.5   & 108.4 & 119.0 &  89.7 &  94.3 & 120.9 & ......\\

      .     &      .   &    .   &    .  &   .   &    .  &     . &     . &     . &    .  &   .   \\
      .     &      .   &    .   &    .  &   .   &    .  &     . &     . &     . &    .  &   .   \\
      .     &      .   &    .   &    .  &   .   &    .  &     . &     . &     . &    .  &   .   \\
\enddata
\tablecomments{Table \ref{tbl-2} is published in its entirety in the
electronic edition of the {\it Astrophysical Journal}.  A portion is
shown here for guidance regarding its form and content.}
\end{deluxetable}

\clearpage
\begin{deluxetable}{lcccccccc}
\tabletypesize{\scriptsize} \tablecolumns{9} \tablewidth{0pc}
\tablecaption{$T_{\mathrm{eff}}$ derived by different methods. }
\tablehead{ \colhead{} & \colhead{} &
\multicolumn{4}{c}{$T_{\mathrm{eff}}$ [K]} & \colhead{}
&\multicolumn{2}{c}{$T_{\rm{eff}}^{\star}$ [K]} \\
\cline{3-6} \cline{8-9}\\
\colhead{star} & \colhead{E(\bv)} & \colhead{(\textit{V-K})}  &
\colhead{(\textit{J-H})} & \colhead{(\textit{J-K})} &
\colhead{Average} & \colhead{} &\colhead{SA II} & \colhead{This
Work}} \startdata
BD+04\degr~2466   & 0.042 & 5165 & 5120 & 5227 & 5170 & & 5096 & 5115 \\
BD+01\degr~3070   & 0.054 & 5352 & 5319 & 5181 & 5284 & & 5132 & 5130 \\
BD+09\degr~2870   & 0.029 &  ... & 4247 & ...  & 4247 & & 4324 & 4300 \\
BD+10\degr~2495   & 0.022 & 5042 & 5047 & 4968 & 5019 & & 4801 & 4775 \\
BD+12\degr~2547   & 0.029 & 4668 & 4580 & 4661 & 4636 & & 4573 & 4590 \\
BD+29\degr~2356   & 0.013 & 4648 & 4676 & 4824 & 4716 & & 4816 & 4795 \\
BD+30\degr~2611   & 0.019 &  ... &  ... & ...  & ...  & & 4280 & 4300 \\
HD~33771          & 0.030 & 4761 & 4717 & 4716 & 4731 & & 4575 & 4635 \\
HD~85773          & 0.047 &  ... & 4250 & ...  & 4250 & & 4263 & 4205 \\
HD~107752         & 0.030 &  ... &  ... & ...  & ...  & & 4385 & 4370 \\
HD~108577         & 0.027 &  ... & 4743 & ...  & 4743 & & 4795 & 4800 \\
HD~119516         & 0.026 & 5506 & 5565 & 5586 & 5552 & & 5512 & 5550 \\
HD~124358         & 0.064 & 4702 & 4677 & 4786 & 4721 & & 4674 & 4660 \\
HD~128279         & 0.047 &  ... & 4918 & ...  & 4918 & & 5059 & 5100 \\
HD~175305         & 0.034 & 5047 & 5068 & 5013 & 5042 & & 4973 & 5035 \\
HD~237846         & 0.010 &  ... & 4977 & 4994 & 4985 & & 4676 & 4725 \\
HD~134439         & 0.011 &  ... &  ... & ...  & ...  & & 5100 & 5080 \\
G~112--43         & 0.012 & 6003 & 6000 & 6046 & 6016 & & 6057 & 6000 \\
G~115--58         & 0.011 & 6126 & 6089 & 6134 & 6116 & & 6050 & 6080 \\
G~15--13          & 0.020 & 5046 & 5522 & 5450 & 5339 & & 5203 & 5125 \\
G~166--37         & 0.024 & 5314 & 5166 & 5335 & 5272 & & 5401 & 5335 \\
G~238--30         & 0.018 & 5372 & 5742 & 5762 & 5625 & & 5553 & 5490 \\
G~41--41          & 0.030 & 6444 & 6111 & 6233 & 6263 & & 6109 & 6100 \\
G~48--29          & 0.058 &  ... & 6345 & 6204 & 6274 & & 6119 & 6200 \\
G~53--41          & 0.040 & 5930 & 6011 & 5931 & 5957 & & 5825 & 5840 \\
LP~894--3         & 0.014 & 5755 & 5849 & 5825 & 5816 & & 5827 & 5860 \\

\enddata
\tablecomments{Column 3 to column 6 represent the values estimated
by color indices. Last two columns represent the values derived by
different spectroscopic analysis codes.}
\end{deluxetable}

\clearpage
\begin{deluxetable}{lclccc}
\tabletypesize{\scriptsize} \tablecaption{Derived atmospheric
parameters for the sample stars} \tablewidth{0pt}
 \tablehead{ \colhead{star} & \colhead{Type} & \colhead{$T_{\mathrm{eff}}$ [K]} &
\colhead{$\log g$} & \colhead{[Fe/H]} &\colhead{$\xi$
[km$\cdot$s$^{-1}$]} } \startdata
BD+04\degr~2466  &  G~& $5115\pm 80$ & $1.87\pm0.23$ & $-1.92\pm0.05$ & $1.70\pm0.2$  \\
BD+01\degr~3070  &  G~& $5130\pm110$ & $3.10\pm0.10$ & $-1.52\pm0.10$ & $1.20\pm0.1$  \\
BD+09\degr~2870  &  G~& $4300\pm100$ & $0.51\pm0.15$ & $-2.73\pm0.13$ & $1.60\pm0.3$  \\
BD+10\degr~2495  &  G~& $4775\pm100$ & $1.90\pm0.15$ & $-2.10\pm0.08$ & $1.47\pm0.3$  \\
BD+12\degr~2547  &  G~& $4590\pm 70$ & $1.50\pm0.15$ & $-1.89\pm0.10$ & $1.50\pm0.2$  \\
BD+29\degr~2356  &  G~& $4795\pm 60$ & $2.03\pm0.23$ & $-1.45\pm0.10$ & $1.40\pm0.2$  \\
BD+30\degr~2611  &  G~& $4300\pm 70$ & $1.25\pm0.10$ & $-1.41\pm0.09$ & $1.80\pm0.2$  \\
HD~33771         &  G~& $4635\pm 50$ & $1.54\pm0.15$ & $-2.06\pm0.07$ & $1.60\pm0.3$  \\
HD~85773         &  G~& $4205\pm 50$ & $0.32\pm0.20$ & $-2.58\pm0.10$ & $1.60\pm0.3$  \\
HD~107752        &  G~& $4370\pm 90$ & $0.54\pm0.20$ & $-3.16\pm0.10$ & $1.58\pm0.3$  \\
HD~108577        &  G~& $4800\pm 80$ & $1.13\pm0.10$ & $-2.37\pm0.09$ & $1.80\pm0.3$  \\
HD~119516        &  G~& $5550\pm 50$ & $2.20\pm0.20$ & $-1.82\pm0.10$ & $2.00\pm0.3$  \\
HD~124358        &  G~& $4660\pm 70$ & $1.25\pm0.20$ & $-1.77\pm0.12$ & $1.80\pm0.2$  \\
HD~128279        &  G~& $5100\pm 50$ & $2.92\pm0.10$ & $-2.24\pm0.10$ & $1.40\pm0.3$  \\
HD~175305        &  G~& $5035\pm 70$ & $2.84\pm0.10$ & $-1.35\pm0.10$ & $1.40\pm0.1$  \\
HD~237846        &  G~& $4725\pm 60$ & $1.41\pm0.10$ & $-3.01\pm0.10$ & $1.50\pm0.3$  \\
HD~134439        &  D & $5080\pm110$ & $4.94\pm0.22$ & $-1.35\pm0.08$ & $0.00\pm0.4$  \\
G~112--43        &  D & $6000\pm 70$ & $4.00\pm0.10$ & $-1.31\pm0.10$ & $1.40\pm0.3$  \\
G~115--58        &  D & $6080\pm105$ & $3.88\pm0.24$ & $-1.36\pm0.12$ & $1.40\pm0.3$  \\
G~15--13         &  D & $5125\pm100$ & $4.94\pm0.31$ & $-1.63\pm0.09$ & $0.00\pm0.3$  \\
G~166--37        &  D & $5335\pm100$ & $4.98\pm0.30$ & $-1.36\pm0.08$ & $0.00\pm0.1$  \\
G~238--30        &  D & $5490\pm 90$ & $3.57\pm0.31$ & $-3.44\pm0.21$ & $1.46\pm0.3$  \\
G~41--41         &  D & $6100\pm 70$ & $3.90\pm0.30$ & $-2.84\pm0.16$ & $1.42\pm0.2$  \\
G~48--29         &  D & $6200\pm 50$ & $3.66\pm0.10$ & $-2.82\pm0.08$ & $1.60\pm0.3$  \\
G~53--41         &  D & $5840\pm 80$ & $4.51\pm0.24$ & $-1.22\pm0.10$ & $1.10\pm0.1$  \\
LP~894--3        &  D & $5860\pm 80$ & $4.30\pm0.10$ & $-1.50\pm0.11$ & $1.30\pm0.1$  \\

\enddata
\end{deluxetable}

\clearpage
\begin{deluxetable}{llccccc}
\tabletypesize{\scriptsize} \tablecaption{Abundance uncertainties
linked to stellar parameters. } \tablewidth{0pt} \tablehead{Star &
Element Ration & $\Delta T_{\rm{eff}}$($\pm100$ K) & $\Delta\log
g$($\pm0.30 )$ & $\Delta$[Fe/H]($\pm0.10 $) & $\Delta\xi$($\pm0.30$
km$\cdot$s$^{-1}$) & Total Error } \startdata
G166-37   & [Na/Fe] & $\mp0.03$ & $\pm0.00$ & $\pm0.00$ & $\pm0.01$ & $\pm0.03$ \\
(dwarf)   & [Mg/Fe] & $\mp0.03$ & $\pm0.00$ & $\pm0.00$ & $\pm0.00$ & $\pm0.03$ \\
          & [Si/Fe] & $\mp0.02$ & $\pm0.02$ & $\pm0.02$ & $\pm0.01$ & $\pm0.04$ \\
          & [Ca/Fe] & $\mp0.02$ & $\pm0.00$ & $\pm0.00$ & $\pm0.00$ & $\pm0.02$ \\
          & [Sc/Fe] & $\mp0.03$ & $\pm0.01$ & $\pm0.03$ & $\pm0.00$ & $\pm0.04$ \\
          & [Ti/Fe] & $\pm0.00$ & $\mp0.03$ & $\mp0.01$ & $\pm0.01$ & $\pm0.03$ \\
          & [Cr/Fe] & $\pm0.00$ & $\pm0.01$ & $\pm0.00$ & $\pm0.01$ & $\pm0.01$ \\
          & [Mn/Fe] & $\mp0.02$ & $\pm0.02$ & $\pm0.03$ & $\pm0.02$ & $\pm0.05$ \\
          & [Ni/Fe] & $\pm0.01$ & $\pm0.02$ & $\pm0.00$ & $\pm0.01$ & $\pm0.02$ \\
          & [Zn/Fe] & $\mp0.00$ & $\pm0.02$ & $\pm0.01$ & $\pm0.00$ & $\pm0.02$ \\
          & [Y /Fe] & $\mp0.01$ & $\pm0.02$ & $\pm0.01$ & $\pm0.00$ & $\pm0.02$ \\
          & [Ba/Fe] & $\mp0.02$ & $\pm0.07$ & $\pm0.00$ & $\pm0.01$ & $\pm0.09$ \\
          & [Fe/H]$_{\rm{I}}$ & $\pm0.07$ & $\mp0.05$ & $\pm0.02$ & $\pm0.00$ & $\pm0.08$\\
          & [Fe/H]$_{\rm{II}}$ & $\pm0.00$ & $\pm0.06$ & $\pm0.03$ & $\pm0.00$ & $\pm0.06$\\
 & & & & & &  \\
\tableline
 & & & & & &  \\
HD 175305 & [Na/Fe] & $\mp0.04$ & $\pm0.01$ & $\pm0.00$ & $\pm0.06$ & $\pm0.07$ \\
(giant)   & [Mg/Fe] & $\pm0.00$ & $\mp0.04$ & $\pm0.00$ & $\pm0.02$ & $\pm0.04$ \\
          & [Si/Fe] & $\mp0.06$ & $\pm0.03$ & $\pm0.00$ & $\pm0.07$ & $\pm0.09$ \\
          & [Ca/Fe] & $\mp0.02$ & $\mp0.04$ & $\pm0.00$ & $\pm0.03$ & $\pm0.05$ \\
          & [Sc/Fe] & $\mp0.04$ & $\pm0.01$ & $\pm0.01$ & $\pm0.00$ & $\pm0.04$ \\
          & [Ti/Fe] & $\pm0.02$ & $\mp0.01$ & $\pm0.00$ & $\pm0.01$ & $\pm0.02$ \\
          & [Cr/Fe] & $\pm0.01$ & $\mp0.02$ & $\mp0.01$ & $\mp0.01$ & $\pm0.03$ \\
          & [Mn/Fe] & $\mp0.02$ & $\pm0.03$ & $\pm0.00$ & $\pm0.00$ & $\pm0.03$ \\
          & [Ni/Fe] & $\mp0.01$ & $\pm0.02$ & $\pm0.01$ & $\pm0.04$ & $\pm0.04$ \\
          & [Zn/Fe] & $\mp0.00$ & $\pm0.01$ & $\pm0.00$ & $\pm0.01$ & $\pm0.01$ \\
          & [Y /Fe] & $\mp0.05$ & $\pm0.06$ & $\pm0.01$ & $\pm0.03$ & $\pm0.08$ \\
          & [Ba/Fe] & $\mp0.03$ & $\pm0.04$ & $\pm0.01$ & $\mp0.01$ & $\pm0.05$ \\
          & [Fe/H]$_{\rm{I}}$ & $\pm0.06$ & $\mp0.01$ & $\mp0.01$ & $\mp0.07$ & $\pm0.09$\\\
          & [Fe/H]$_{\rm{II}}$ & $\pm0.01$ & $\pm0.09$ & $\pm0.03$ & $\mp0.09$ & $\pm0.11$\\
\enddata
\end{deluxetable}

\clearpage
\begin{deluxetable}{lrrrr}
\tabletypesize{\scriptsize} \tablecaption{Comparisons of stellar
parameters between the two analyses} \tablewidth{0pt}
 \tablehead{ \colhead{star} & \colhead{$\Delta T_{\mathrm{eff}}$} & \colhead{$\Delta \log g$} & \colhead{$\Delta$ [Fe/H]} &\colhead{$\Delta \xi$}}
 \startdata
BD+04\degr~2466  &    19  & $-$0.15  &    0.15 & $-$0.04 \\
BD+01\degr~3070  & $-$2   &    0.03  &    0.06 &    0.10 \\
BD+09\degr~2870  & $-$24  &    0.12  & $-$0.07 & $-$0.33 \\
BD+10\degr~2495  & $-$26  &    0.02  &    0.05 & $-$0.07 \\
BD+12\degr~2547  &    17  &    0.21  &    0.08 & $-$0.20 \\
BD+29\degr~2356  & $-$21  & $-$0.11  & $-$0.01 &    0.00 \\
BD+30\degr~2611  &    20  &    0.51  & $-$0.06 &    0.14 \\
HD~33771         &    60  &    0.18  &    0.07 & $-$0.01 \\
HD~85773         & $-$58  &    0.18  & $-$0.12 & $-$0.41 \\
HD~107752        & $-$15  &    0.10  & $-$0.04 & $-$0.37 \\
HD~108577        &     5  &    0.29  &    0.07 & $-$0.26 \\
HD~119516        &    38  &    0.41  &    0.16 & $-$0.21 \\
HD~124358        & $-$14  &    0.07  & $-$0.01 & $-$0.04 \\
HD~128279        &    41  &    0.29  &    0.14 &    0.06 \\
HD~175305        &    62  &    0.31  &    0.05 &    0.20 \\
HD~237846        &    49  &    0.22  &    0.12 & $-$0.19 \\
HD~134439        & $-$20  & $-$0.05  & $-$0.14 & $-$0.04 \\
G~112--43        & $-$57  &    0.16  &    0.07 &    0.14 \\
G~115--58        &    30  &    0.21  &    0.16 & $-$0.07 \\
G~15--13         & $-$78  & $-$0.06  & $-$0.17 & $-$0.04 \\
G~166--37        & $-$66  & $-$0.02  & $-$0.15 & $-$0.04 \\
G~238--30        & $-$63  & $-$0.32  &    0.08 & $-$0.15 \\
G~41--41         & $-$9   &    0.50  &    0.16 & $-$0.27 \\
G~48--29         &    81  &    0.10  &    0.21 & $-$0.13 \\
G~53--41         &    15  &    0.33  &    0.12 &    0.13 \\
LP~894--3        &    33  &    0.28  &    0.12 &    0.20 \\
 & & & & \\
\tableline
 & & & & \\
$\overline{\Delta}$ & 0.65 & 0.15 & 0.04 & $-$0.07 \\
$\sigma$            &   43 & 0.20 & 0.11 &  0.17 \\
\enddata
\tablecomments{$\Delta$ = SA $-$ SA II }
\end{deluxetable}

\clearpage
\begin{deluxetable}{lrrrrrrrrrr}
\rotate \centering \tabletypesize{\scriptsize}
\tablecaption{Comparisons of abundance results between the two
analyses} \tablewidth{0pt} \tablecolumns{13} \tablehead{
\colhead{Star} & \colhead{} & \colhead{$\Delta$[Si/Fe]} & \colhead{}
& \colhead{$\Delta$[Ti/Fe]} & \colhead{} & \colhead{$\Delta$[Mn/Fe]}
& \colhead{} &
\colhead{$\Delta$[Zn/Fe]} & \colhead{} & \colhead{$\Delta$[Ba/Fe]}\\
\colhead{} & \colhead{$\Delta$[Mg/Fe]}& \colhead{} &
\colhead{$\Delta$[Ca/Fe]} & \colhead{} & \colhead{$\Delta$[Cr/Fe]} &
\colhead{} &  \colhead{$\Delta$[Ni/Fe]} & \colhead{} &
\colhead{$\Delta$[Y/Fe]} & \colhead{}}
 \startdata
BD+04\degr~2466  & $-$0.07  & $-$0.01 & $-$0.01 & $-$0.03 &    0.07 &    0.13 & $-$0.08 & $-$0.06 & $-$0.02 & $-$0.09 \\
BD+01\degr~3070  &    0.10  &    0.14 &    0.02 & $-$0.08 & $-$0.01 &    0.01 & $-$0.02 &    0.04 &    ..   &    0.05 \\
BD+09\degr~2870  & $-$0.01  &    ..   &    0.07 &    0.10 &    0.04 &    0.03 &    0.02 & $-$0.02 &    0.05 &    0.14 \\
BD+10\degr~2495  & $-$0.11  &    0.12 &    0.05 & $-$0.03 & $-$0.11 &    0.08 & $-$0.01 &    0.03 & $-$0.01 &    0.10 \\
BD+12\degr~2547  & $-$0.02  &    0.01 & $-$0.05 & $-$0.12 & $-$0.05 & $-$0.11 &    0.07 &    0.02 &    0.07 & $-$0.05 \\
BD+29\degr~2356  &    0.12  &    0.08 &    0.02 & $-$0.12 &    0.00 & $-$0.20 &    0.01 & $-$0.03 & $-$0.08 & $-$0.12 \\
BD+30\degr~2611  & $-$0.14  &    0.12 & $-$0.04 &    0.01 &    0.10 & $-$0.34 &    0.15 &    0.10 & $-$0.09 & $-$0.03 \\
HD~33771         & $-$0.14  &    0.10 & $-$0.01 &    0.02 & $-$0.10 &    0.03 &    0.04 & $-$0.07 &    0.03 &    0.12 \\
HD~85773         &    0.04  &    ..   &    0.06 &    0.18 & $-$0.02 & $-$0.12 & $-$0.01 &    0.07 &    0.14 &    0.03 \\
HD~107752        &    0.05  &    ..   &    0.09 &    0.12 &    0.09 &    ..   &    0.10 &    0.01 &    0.02 & $-$0.04 \\
HD~108577        &    0.03  &    ..   &    0.01 &    0.07 &    0.05 &    0.12 & $-$0.01 & $-$0.01 &    0.09 &    0.14 \\
HD~119516        &    0.02  &    ..   & $-$0.08 &    0.14 &    0.04 &    0.11 &    0.09 &    0.04 &    0.08 &    0.08 \\
HD~124358        &    ..    &    0.15 & $-$0.01 & $-$0.08 & $-$0.05 &    0.07 &    0.03 &    0.01 &    0.02 & $-$0.11 \\
HD~128279        &    0.07  &    0.09 &    0.00 &    0.05 &    0.04 &    0.17 &    0.05 &    0.01 &    ..   &    0.04 \\
HD~175305        &    0.13  &    0.11 & $-$0.01 & $-$0.09 & $-$0.01 & $-$0.11 & $-$0.01 &    0.07 &    0.06 & $-$0.10 \\
HD~237846        &    0.10  &    ..   & $-$0.01 &    0.08 &    0.16 &    ..   &    ..   & $-$0.03 &    ..   &    0.08 \\
HD~134439        & $-$0.06  & $-$0.12 & $-$0.08 & $-$0.02 & $-$0.02 & $-$0.28 & $-$0.12 &    ..   &    ..   & $-$0.18 \\
G~112--43        &    0.10  &    0.06 &    0.02 &    0.03 & $-$0.01 &    0.11 &    0.01 &    0.01 &    0.06 & $-$0.11 \\
G~115--58        &    0.08  &    0.04 &    0.00 &    0.07 &    0.05 &    0.03 &    0.02 &    0.03 &    0.15 & $-$0.14 \\
G~15--13         &    ..    &    ..   &    ..   &    0.04 &    0.05 & $-$0.21 &    0.01 &    0.20 & $-$0.18 & $-$0.16 \\
G~166--37        & $-$0.11  &    0.01 &    ..   &    0.01 &    0.05 & $-$0.34 &    0.02 & $-$0.09 & $-$0.20 & $-$0.10 \\
G~238--30        & $-$0.06  &    ..   &    0.05 &    0.11 &    0.17 &    ..   &    ..   &    ..   &    ..   &   ..    \\
G~41--41         & $-$0.07  &    ..   & $-$0.03 &    0.13 &    ..   &    ..   &    ..   &    ..   &    ..   &    0.15 \\
G~48--29         &  0.01    &    ..   & $-$0.01 & $-$0.04 &    0.00 &    ..   &    ..   &    ..   &    ..   & $-$0.01 \\
G~53--41         &  0.04    & $-$0.08 & $-$0.04 &    0.03 & $-$0.07 &    0.06 &    0.02 & $-$0.01 & $-$0.02 &    0.00 \\
LP~894--3        & $-$0.06  &    0.08 &    0.01 &    0.10 &    0.09 &    ..   &    0.08 &    0.05 &    ..   & $-$0.14 \\
 & & & & & & & & & & \\
\tableline
 & & & & & & & & & & \\
$\overline{\Delta\rm [X/Fe]}$ & 0.00 & 0.06 & 0.00 & 0.03 & 0.02 & $-$0.04 & 0.02 & 0.02 & 0.01 & $-$0.02 \\
 $\sigma$                     & 0.08 & 0.08 & 0.04 & 0.08 & 0.07 &    0.14 & 0.06 & 0.07 & 0.10 &    0.11 \\
\enddata
\tablecomments{$\Delta$ = SA $-$ SA II. $\Delta$[Na/Fe] is not
presented here because of no non-LTE correction adopted in SA II. }
\end{deluxetable}

\clearpage
\begin{deluxetable}{lcccccccccccccc}
\tabletypesize{\scriptsize} \tablecolumns{14} \tablewidth{0pc}
\tablecaption{Comparison of parameters of common stars with SB02  }
\tablehead{
 \colhead{} & \multicolumn{4}{c}{Parameters (TW)} & \colhead{}
&\multicolumn{4}{c}{Parameters (SB02)} & \colhead{}
&\multicolumn{4}{c}{$\Delta$ (TW-SB02)}\\
\cline{2-5} \cline{7-10} \cline{12-15}\\
\colhead{star}  & \colhead{$T_{\mathrm{eff}}$} & \colhead{$\log g$}
& \colhead{[Fe/H]} & \colhead{$\xi$} & \colhead{} &
\colhead{$T_{\mathrm{eff}}$} & \colhead{$\log g$} & \colhead{[Fe/H]}
& \colhead{$\xi$} & \colhead{} & \colhead{$\Delta T_{\mathrm{eff}}$}
& \colhead{$\Delta \log g$} & \colhead{$\Delta$ [Fe/H]} &
\colhead{$\Delta \xi$} } \startdata
G~15-13  & $5125$ & $4.94$ & $-1.63$ & $0.00$ & & 5082 & 4.61 & $-$1.70 & 0.00 & &    43 & 0.33 & 0.07 & 0.00\\
G~166-37 & $5335$ & $4.98$ & $-1.36$ & $0.00$ & & 5350 & 4.71 & $-$1.39 & 0.00 & & $-$15 & 0.27 & 0.03 & 0.00\\
G~238-30 & $5490$ & $3.57$ & $-3.44$ & $1.46$ & & 5383 & 3.43 & $-$3.60 & 1.19 & &   107 & 0.15 & 0.16 & 0.27\\
\enddata
\tablecomments{Column 2 $\sim$ column 5 are the results derived in
this work (TW), the results of column 6 $\sim$ column 9 are cited
from SB02, and the last 4 columns represent the discrepancy of these
two works.}
\end{deluxetable}

\clearpage
\begin{deluxetable}{lrrrrrrrrrr}
\centering \tabletypesize{\scriptsize} \tablecaption{Comparison of
abundances with SB02 } \tablewidth{0pt} \tablecolumns{11}
\tablehead{ \colhead{Star} & \colhead{$\Delta$[Fe/H]} & \colhead{} &
\colhead{$\Delta$[Mg/Fe]} & \colhead{} & \colhead{$\Delta$[Ca/Fe]} &
\colhead{} & \colhead{$\Delta$[Cr/Fe]}
& \colhead{} & \colhead{$\Delta$[Y/Fe]} &  \colhead{}\\
\colhead{} & \colhead{} & \colhead{$\Delta$[Na/Fe]}& \colhead{} &
\colhead{$\Delta$[Si/Fe]} & \colhead{} & \colhead{$\Delta$[Ti/Fe]} &
\colhead{} &  \colhead{$\Delta$[Ni/Fe]} & \colhead{} &
\colhead{$\Delta$[Ba/Fe]}}
 \startdata
G~15--13$^{a}$   & 0.07 &    0.23 &    0.05 &    ...  & $-$0.11 &    0.08 & 0.08 &    0.08 &    0.00 &    0.07 \\
G~166--37$^{a}$  & 0.03 &    0.13 &    0.04 &    0.02 & $-$0.05 &    0.08 &-0.10 &    0.09 & $-$0.03 &    0.02 \\
G~238--30$^{a}$  & 0.16 &    ...  &    0.12 &    ...  &    0.01 &    0.25 &-0.09 &    ...  &    ...  &    ...  \\
 & & & & & & & & &  & \\
\tableline
 & & & & & & & & &  & \\
G~5--19$^{b}$    & 0.02 & $-$0.02 & $-$0.08 &    0.05 & $-$0.05 & $-$0.06 & 0.06 & $-$0.05 &    0.03 & $-$0.07 \\
G~9--36$^{b}$    & 0.03 &    0.01 & $-$0.04 &    0.16 &    0.02 & $-$0.01 & 0.07 &    0.08 & $-$0.08 & $-$0.06 \\
G~215--47$^{b}$  & 0.00 & $-$0.03 & $-$0.11 & $-$0.13 & $-$0.09 & $-$0.09 & 0.03 & $-$0.16 &    0.09 & $-$0.03 \\
 & & & & & & & & &  & \\
\tableline
 & & & & & & & & &  & \\
G~5--19$^{c}$    & 0.00 & $-$0.02 & $-$0.05 &    0.03 & $-$0.02 & $-$0.02 &-0.01 &    0.00 & $-$0.01 & $-$0.05 \\
G~9--36$^{c}$    & 0.00 &    0.00 & $-$0.02 & $-$0.05 &    0.02 &    0.02 & 0.03 & $-$0.06 & $-$0.04 & $-$0.04 \\
G~215--47$^{c}$  & 0.00 & $-$0.02 & $-$0.07 &    0.02 & $-$0.03 &    0.00 &-0.03 & $-$0.03 &    0.02 & $-$0.03 \\
\enddata
\tablecomments{$\Delta$ = TW $-$ SB02; a: Abundances derived from our measured equivalent widths;
    b \& c: Abundance derived from the equivalent widths presented by SB02:  abundances of b were determined taking
    atmospheric parameters derived in this work, while abundances of c were determined taking parameters presented in SB02. }
\end{deluxetable}

\clearpage
\begin{deluxetable}{lrrrrrrrrrr}
\tabletypesize{\scriptsize}
 \rotate
 \tablecaption{Abundance ratios: Na, Mg, Si, and Ca}
 \tablewidth{0pt}
 \tablehead{
\colhead{Star} & \colhead{[NaI/H]} & \colhead{N} & \colhead{[MgI/H]}
& \colhead{N} & \colhead{[SiI/H]} & \colhead{N} & \colhead{[SiII/H]}
& \colhead{N} & \colhead{[CaI/H]}& \colhead{N}}
 \startdata
BD+04\degr 2466 & $-1.76\pm0.08$ & 2 & $-1.67\pm0.08$ & 1 & $-1.42\pm0.07$ & 2 & $    ...     $ & . & $-1.47\pm0.08$ & 21 \\
BD+01\degr 3070 & $-1.61\pm0.06$ & 3 & $-1.20\pm0.09$ & 3 & $-1.16\pm0.09$ & 3 & $-1.17\pm0.06$ & 1 & $-1.28\pm0.06$ & 19 \\
BD+09\degr 2870 & $-2.62\pm0.09$ & 2 & $-2.26\pm0.13$ & 2 & $    ...     $ & . & $    ...     $ & . & $-2.26\pm0.10$ & 25 \\
BD+10\degr 2495 & $-2.04\pm0.06$ & 2 & $-1.88\pm0.08$ & 1 & $-1.65\pm0.05$ & 3 & $    ...     $ & . & $-1.76\pm0.06$ & 23 \\
BD+12\degr 2547 & $-2.04\pm0.06$ & 1 & $-1.53\pm0.08$ & 1 & $-1.59\pm0.09$ & 2 & $    ...     $ & . & $-1.66\pm0.06$ & 21 \\
BD+29\degr 2356 & $-1.49\pm0.09$ & 3 & $-1.09\pm0.08$ & 1 & $-1.10\pm0.07$ & 3 & $-1.20\pm0.04$ & 1 & $-1.20\pm0.07$ & 18 \\
BD+30\degr 2611 & $-1.74\pm0.08$ & 2 & $-1.27\pm0.08$ & 1 & $-1.17\pm0.08$ & 3 & $-1.07\pm0.05$ & 1 & $-1.36\pm0.07$ & 12 \\
HD     33771    & $-2.10\pm0.06$ & 2 & $-1.77\pm0.08$ & 1 & $-1.62\pm0.07$ & 2 & $    ...     $ & . & $-1.76\pm0.06$ & 19 \\
HD     85773    & $-3.29\pm0.08$ & 1 & $-2.29\pm0.09$ & 3 & $    ...     $ & . & $    ...     $ & . & $-2.34\pm0.07$ & 17 \\
HD     107752   & $-3.63\pm0.05$ & 1 & $-2.78\pm0.10$ & 2 & $    ...     $ & . & $    ...     $ & . & $-2.62\pm0.06$ & 15 \\
HD     108577   & $-2.32\pm0.05$ & 2 & $-1.95\pm0.08$ & 2 & $    ...     $ & . & $    ...     $ & . & $-1.97\pm0.06$ & 20 \\
HD     119516   & $-1.66\pm0.07$ & 2 & $-1.66\pm0.09$ & 3 & $    ...     $ & . & $    ...     $ & . & $-1.60\pm0.07$ & 19 \\
HD     124358   & $-2.05\pm0.06$ & 2 & $    ...     $ & . & $-1.43\pm0.05$ & 3 & $    ...     $ & . & $-1.62\pm0.07$ & 19 \\
HD     128279   & $-2.62\pm0.09$ & 2 & $-1.80\pm0.07$ & 2 & $-1.77\pm0.08$ & 2 & $    ...     $ & . & $-1.89\pm0.07$ & 20 \\
HD     175305   & $-1.37\pm0.07$ & 2 & $-1.01\pm0.08$ & 1 & $-1.04\pm0.10$ & 3 & $-0.97\pm0.05$ & 1 & $-1.12\pm0.07$ & 19 \\
HD     237846   & $-3.40\pm0.05$ & 1 & $-2.54\pm0.10$ & 2 & $    ...     $ & . & $    ...     $ & . & $-2.64\pm0.08$ & 14 \\
HD     134439   & $-1.79\pm0.06$ & 1 & $-1.27\pm0.12$ & 2 & $-1.30\pm0.05$ & 3 & $    ...     $ & . & $-1.34\pm0.05$ & 19 \\
G 112-43        & $-1.34\pm0.06$ & 1 & $-0.96\pm0.12$ & 1 & $-1.05\pm0.06$ & 3 & $-1.05\pm0.05$ & 1 & $-1.03\pm0.06$ & 23 \\
G 115-58        & $-1.79\pm0.05$ & 3 & $-1.21\pm0.08$ & 2 & $-1.25\pm0.07$ & 2 & $    ...     $ & . & $-1.13\pm0.07$ & 14 \\
G 15-13         & $-2.00\pm0.06$ & 2 & $-1.42\pm0.08$ & 1 & $    ...     $ & . & $    ...     $ & . & $-1.54\pm0.07$ & 19 \\
G 166-37        & $-1.58\pm0.05$ & 2 & $-1.11\pm0.07$ & 2 & $-1.18\pm0.07$ & 2 & $    ...     $ & . & $-1.21\pm0.05$ & 20 \\
G 238-30        & $-4.09\pm0.06$ & 2 & $-2.70\pm0.08$ & 1 & $    ...     $ & . & $    ...     $ & . & $-2.75\pm0.06$ &  3 \\
G 41-41         & $-3.55\pm0.07$ & 2 & $-2.39\pm0.08$ & 1 & $    ...     $ & . & $    ...     $ & . & $-2.38\pm0.06$ &  2 \\
G 48-29         & $-3.32\pm0.06$ & 2 & $-2.26\pm0.08$ & 1 & $    ...     $ & . & $    ...     $ & . & $-2.24\pm0.09$ &  5 \\
G 53-41         & $-0.97\pm0.08$ & 2 & $-1.00\pm0.08$ & 1 & $-0.99\pm0.07$ & 2 & $-0.89\pm0.04$ & 1 & $-1.02\pm0.06$ & 22 \\
LP 894-3        & $-1.67\pm0.06$ & 2 & $-1.36\pm0.08$ & 2 & $-1.22\pm0.05$ & 1 & $    ...     $ & . & $-1.36\pm0.06$ & 20 \\
\enddata
\end{deluxetable}

\clearpage
\begin{deluxetable}{lrrrrrrrrrrrr}
\tabletypesize{\scriptsize}
 \rotate
 \tablecaption{Abundance ratios: Sc, Ti, Cr and Mn} \tablewidth{0pt}
 \tablehead{
    \colhead{star} & \colhead{[ScII/H]}  & \colhead{N} & \colhead{[TiI/H]}
& \colhead{N} & \colhead{[TiII/H]} & \colhead{N} & \colhead{[CrI/H]}
& \colhead{N} & \colhead{[CrII/H]} & \colhead{N} & \colhead{[MnI/H]}
& \colhead{N}}
 \startdata
BD+04\degr 2466 & $-1.75\pm0.08$ & 4& $-1.55\pm0.08$ & 33 & $-1.55\pm0.08$ &  8& $-2.02\pm0.08$ & 14 & $-1.80\pm0.09$ & 5 & $-2.36\pm0.04$ & 2 \\
BD+01\degr 3070 & $-1.29\pm0.05$ & 7& $-1.29\pm0.06$ & 32 & $-1.11\pm0.08$ &  8& $-1.66\pm0.06$ & 14 & $-1.37\pm0.09$ & 5 & $-2.10\pm0.03$ & 2 \\
BD+09\degr 2870 & $-2.69\pm0.05$ & 6& $-2.50\pm0.11$ & 32 & $-2.44\pm0.10$ &  9& $-3.05\pm0.05$ & 12 & $-2.59\pm0.09$ & 5 & $-3.34\pm0.07$ & 2 \\
BD+10\degr 2495 & $-1.98\pm0.06$ & 9& $-1.93\pm0.06$ & 32 & $-1.78\pm0.08$ &  7& $-2.28\pm0.09$ & 14 & $-1.91\pm0.07$ & 5 & $-2.59\pm0.03$ & 2 \\
BD+12\degr 2547 & $-1.69\pm0.07$ & 3& $-1.79\pm0.06$ & 29 & $-1.50\pm0.09$ &  7& $-2.09\pm0.07$ & 15 & $-1.74\pm0.09$ & 5 & $-2.49\pm0.03$ & 2 \\
BD+29\degr 2356 & $-1.20\pm0.07$ & 2& $-1.21\pm0.06$ & 27 & $-1.00\pm0.10$ &  6& $-1.57\pm0.07$ & 14 & $-1.31\pm0.11$ & 5 & $-2.12\pm0.08$ & 2 \\
BD+30\degr 2611 & $    ...     $ & .& $-1.36\pm0.07$ &  9 & $-1.13\pm0.08$ &  2& $-1.43\pm0.10$ &  3 & $-1.35\pm0.09$ & 2 & $-2.06\pm0.04$ & 2 \\
HD 33771        & $-1.90\pm0.08$ & 4& $-1.87\pm0.05$ & 31 & $-1.68\pm0.08$ &  8& $-2.30\pm0.05$ & 12 & $-1.92\pm0.09$ & 5 & $-2.60\pm0.04$ & 2 \\
HD 85773        & $-2.60\pm0.06$ & 4& $-2.30\pm0.07$ & 21 & $-2.25\pm0.08$ &  7& $-2.85\pm0.07$ & 12 & $-2.40\pm0.10$ & 4 & $-3.02\pm0.06$ & 2 \\
HD 107752       & $-3.30\pm0.08$ & 7& $-3.03\pm0.07$ & 14 & $-2.98\pm0.07$ & 11& $-3.39\pm0.07$ & 10 & $-3.02\pm0.06$ & 2 & $-3.71\pm0.03$ & 1 \\
HD 108577       & $-2.24\pm0.07$ &10& $-2.10\pm0.05$ & 25 & $-2.06\pm0.08$ &  8& $-2.53\pm0.07$ & 14 & $-2.28\pm0.11$ & 3 & $-2.68\pm0.04$ & 2 \\
HD 119516       & $-1.74\pm0.07$ & 8& $-1.51\pm0.10$ & 20 & $-1.55\pm0.08$ &  8& $-1.97\pm0.08$ & 12 & $-1.74\pm0.08$ & 5 & $-2.23\pm0.04$ & 2 \\
HD 124358       & $-1.73\pm0.07$ & 3& $-1.73\pm0.06$ & 27 & $-1.54\pm0.08$ &  5& $-2.01\pm0.06$ & 12 & $-1.69\pm0.11$ & 4 & $-2.21\pm0.03$ & 1 \\
HD 128279       & $-2.02\pm0.07$ & 9& $-1.94\pm0.05$ & 21 & $-1.80\pm0.07$ & 11& $-2.42\pm0.05$ & 12 & $-2.04\pm0.09$ & 3 & $-2.66\pm0.03$ & 2 \\
HD 175305       & $-1.00\pm0.10$ & 5& $-1.11\pm0.06$ & 30 & $-0.90\pm0.08$ &  8& $-1.47\pm0.07$ & 13 & $-1.16\pm0.09$ & 5 & $-1.94\pm0.05$ & 2 \\
HD 237846       & $-3.05\pm0.09$ & 7& $-2.74\pm0.06$ & 19 & $-2.73\pm0.07$ & 11& $-3.25\pm0.06$ & 10 & $-3.04\pm0.04$ & 1 & $-3.49\pm0.03$ & 2 \\
HD 134439       & $-1.27\pm0.09$ & 6& $-1.10\pm0.06$ & 26 & $-1.01\pm0.08$ &  8& $-1.34\pm0.07$ & 15 & $-1.12\pm0.09$ & 5 & $-1.86\pm0.03$ & 1 \\
G 112-43        & $-1.07\pm0.05$ & 9& $-0.91\pm0.05$ & 27 & $-0.86\pm0.07$ &  2& $-1.37\pm0.05$ & 13 & $-1.20\pm0.04$ & 2 & $-1.58\pm0.03$ & 2 \\
G 115-58        & $-1.31\pm0.07$ & 8& $-1.06\pm0.06$ & 18 & $-1.01\pm0.08$ & 10& $-1.47\pm0.05$ & 11 & $-1.32\pm0.10$ & 4 & $-1.78\pm0.03$ & 2 \\
G 15-13         & $-1.48\pm0.08$ & 8& $-1.34\pm0.06$ & 27 & $-1.25\pm0.06$ &  9& $-1.69\pm0.06$ & 19 & $-1.37\pm0.04$ & 1 & $-2.03\pm0.10$ & 2 \\
G 166-37        & $-1.02\pm0.06$ & 8& $-1.05\pm0.05$ & 26 & $-0.93\pm0.08$ & 10& $-1.42\pm0.06$ & 15 & $-1.21\pm0.04$ & 2 & $-1.84\pm0.06$ & 1 \\
G 238-30        & $-3.40\pm0.20$ & 4& $-2.95\pm0.10$ &  3 & $-3.20\pm0.09$ &  3& $-3.57\pm0.09$ &  2 & $    ...     $ & . & $    ...     $ & . \\
G 41-41         & $-2.50\pm0.10$ & 6& $    ...     $ &  . & $-2.25\pm0.09$ &  3& $    ...     $ &  . & $    ...     $ & . & $    ...     $ & . \\
G 48-29         & $-2.56\pm0.08$ & 4& $    ...     $ &  . & $-2.26\pm0.10$ &  3& $-2.89\pm0.10$ &  2 & $    ...     $ & . & $    ...     $ & . \\
G 53-41         & $-1.04\pm0.07$ & 8& $-0.96\pm0.08$ & 29 & $-0.80\pm0.08$ & 11& $-1.35\pm0.06$ & 14 & $-1.03\pm0.09$ & 5 & $-1.62\pm0.03$ & 2 \\
LP 894-3        & $-1.45\pm0.09$ & 9& $-1.32\pm0.11$ & 13 & $-1.19\pm0.08$ & 10& $-1.65\pm0.05$ & 11 & $-1.29\pm0.10$ & 4 & $-1.95\pm0.03$ & 1 \\
\enddata
\end{deluxetable}

\clearpage
\begin{deluxetable}{lcrrrrrrrrrrr}
\tabletypesize{\scriptsize} \rotate \tablecaption{Abundance ratios:
Fe, Ni, Zn, Y and Ba}
 \tablewidth{0pt}
 \tablehead{
\colhead{Star} & \colhead{[FeI/H]} & \colhead{N} &
\colhead{[FeII/H]} & \colhead{N} & \colhead{[NiI/H]} & \colhead{N} &
\colhead{[ZnI/H]} & \colhead{N} & \colhead{[YII/H]} & \colhead{N} &
\colhead{[BaII/H]} & \colhead{N}} \startdata
BD+04\degr 2466 & $-1.92\pm0.05$ &  80 & $-1.92\pm0.09$ & 12 & $-1.90\pm0.08$ & 15 & $-1.96\pm0.10$ & 3& $-1.37\pm0.05$ & 1 & $-0.61\pm0.05$ & 3 \\
BD+01\degr 3070 & $-1.52\pm0.05$ &  91 & $-1.52\pm0.06$ & 12 & $-1.63\pm0.07$ & 22 & $-1.58\pm0.03$ & 3& $-1.66\pm0.05$ & 2 & $-1.59\pm0.06$ & 3 \\
BD+09\degr 2870 & $-2.73\pm0.05$ &  85 & $-2.73\pm0.06$ & 12 & $-2.74\pm0.07$ & 15 & $-2.78\pm0.03$ & 2& $-3.21\pm0.05$ & 2 & $-3.53\pm0.06$ & 3 \\
BD+10\degr 2495 & $-2.10\pm0.05$ &  89 & $-2.10\pm0.06$ & 12 & $-2.17\pm0.07$ & 21 & $-2.11\pm0.03$ & 3& $-2.36\pm0.05$ & 2 & $-2.21\pm0.06$ & 3 \\
BD+12\degr 2547 & $-1.89\pm0.05$ &  77 & $-1.89\pm0.06$ &  9 & $-1.83\pm0.07$ & 22 & $-1.96\pm0.03$ & 2& $-2.03\pm0.05$ & 2 & $-1.99\pm0.06$ & 3 \\
BD+29\degr 2356 & $-1.45\pm0.05$ &  74 & $-1.46\pm0.06$ & 10 & $-1.54\pm0.08$ & 22 & $-1.53\pm0.03$ & 3& $-1.5 \pm0.05$ & 2 & $-1.44\pm0.06$ & 3 \\
BD+30\degr 2611 & $-1.40\pm0.06$ &  55 & $-1.41\pm0.08$ &  7 & $-1.56\pm0.09$ & 15 & $-1.73\pm0.03$ & 2& $-1.63\pm0.05$ & 1 & $-1.46\pm0.06$ & 3 \\
HD 33771        & $-2.06\pm0.05$ &  85 & $-2.06\pm0.06$ & 10 & $-2.13\pm0.07$ & 16 & $-2.14\pm0.04$ & 3& $-2.21\pm0.05$ & 1 & $-2.15\pm0.06$ & 3 \\
HD 85773        & $-2.58\pm0.05$ &  60 & $-2.58\pm0.05$ &  9 & $-2.58\pm0.08$ & 15 & $-2.14\pm0.05$ & 3& $-3.15\pm0.05$ & 1 & $-3.27\pm0.06$ & 4 \\
HD 107752       & $-3.16\pm0.05$ &  84 & $-3.16\pm0.05$ & 12 & $-3.18\pm0.08$ &  9 & $-3.11\pm0.04$ & 3& $-3.56\pm0.06$ & 2 & $-3.72\pm0.06$ & 4 \\
HD 108577       & $-2.37\pm0.05$ &  87 & $-2.37\pm0.07$ & 12 & $-2.29\pm0.07$ & 15 & $-2.27\pm0.04$ & 2& $-2.69\pm0.08$ & 2 & $-2.54\pm0.06$ & 3 \\
HD 119516       & $-1.82\pm0.05$ &  82 & $-1.82\pm0.07$ & 10 & $-1.89\pm0.10$ &  6 & $-1.96\pm0.07$ & 3& $-1.96\pm0.05$ & 2 & $-1.71\pm0.06$ & 3 \\
HD 124358       & $-1.77\pm0.05$ &  68 & $-1.77\pm0.06$ &  9 & $-1.94\pm0.07$ & 23 & $-2.05\pm0.04$ & 6& $-2.37\pm0.05$ & 1 & $-2.20\pm0.06$ & 3 \\
HD 128279       & $-2.24\pm0.05$ &  85 & $-2.24\pm0.06$ & 11 & $-2.21\pm0.08$ & 11 & $-2.27\pm0.06$ & 3& $    ...     $ & . & $-2.97\pm0.07$ & 3 \\
HD 175305       & $-1.35\pm0.09$ &  80 & $-1.35\pm0.11$ & 11 & $-1.44\pm0.07$ & 22 & $-1.38\pm0.04$ & 3& $-1.35\pm0.09$ & 2 & $-1.36\pm0.06$ & 3 \\
HD 237846       & $-3.01\pm0.05$ &  77 & $-3.01\pm0.06$ & 12 & $-3.06\pm0.12$ &  5 & $-2.95\pm0.04$ & 2& $-3.52\pm0.07$ & 2 & $-4.09\pm0.10$ & 3 \\
HD 134439       & $-1.35\pm0.05$ &  70 & $-1.35\pm0.08$ & 11 & $-1.47\pm0.08$ & 21 & $-1.45\pm0.03$ & 2& $-1.65\pm0.05$ & 1 & $-1.59\pm0.06$ & 4 \\
G 112-43        & $-1.31\pm0.05$ &  95 & $-1.31\pm0.06$ & 13 & $-1.29\pm0.05$ & 14 & $-1.12\pm0.04$ & 3& $-1.4 \pm0.05$ & 2 & $-1.61\pm0.06$ & 3 \\
G 115-58        & $-1.36\pm0.05$ &  71 & $-1.36\pm0.06$ &  9 & $-1.49\pm0.10$ &  8 & $-1.57\pm0.07$ & 2& $-1.44\pm0.05$ & 2 & $-1.51\pm0.10$ & 4 \\
G 15-13         & $-1.63\pm0.05$ &  78 & $-1.63\pm0.08$ & 11 & $-1.71\pm0.08$ & 19 & $-1.65\pm0.04$ & 2& $-1.66\pm0.06$ & 2 & $-1.41\pm0.09$ & 4 \\
G 166-37        & $-1.36\pm0.08$ &  86 & $-1.36\pm0.07$ & 12 & $-1.43\pm0.03$ & 18 & $-1.62\pm0.04$ & 2& $-1.27\pm0.06$ & 2 & $-1.19\pm0.10$ & 4 \\
G 238-30        & $-3.44\pm0.06$ &  20 & $-3.44\pm0.09$ &  3 & $    ...     $ &  . & $    ...     $ & .& $    ...     $ & . & $    ...     $ & . \\
G 41-41         & $-2.81\pm0.06$ &  50 & $-2.82\pm0.07$ &  6 & $    ...     $ &  . & $    ...     $ & .& $    ...     $ & . & $-3.20\pm0.06$ & 1 \\
G 48-29         & $-2.63\pm0.05$ &  24 & $-2.64\pm0.07$ &  7 & $    ...     $ &  . & $    ...     $ & .& $    ...     $ & . & $-3.36\pm0.08$ & 2 \\
G 53-41         & $-1.22\pm0.05$ &  93 & $-1.22\pm0.06$ & 12 & $-1.29\pm0.07$ & 17 & $-1.28\pm0.05$ & 3& $-1.12\pm0.05$ & 2 & $-0.95\pm0.06$ & 3 \\
LP 894-3        & $-1.50\pm0.05$ &  82 & $-1.50\pm0.05$ & 11 & $-1.59\pm0.10$ & 10 & $-1.74\pm0.03$ & 2& $    ...     $ & . & $-1.70\pm0.06$ & 4 \\
\enddata
\end{deluxetable}

\clearpage
\begin{deluxetable}{lcrrrrrr}
\tabletypesize{\scriptsize} \rotate \tablecaption{Abundance relative
to Fe: Na, Mg, Si, Ca, Sc, and Ti } \tablewidth{0pt}
\tablehead{\colhead{Star} & \colhead{[Fe/H ]} & \colhead{[Na/Fe]} &
\colhead{[Mg/Fe]} & \colhead{[Si/Fe]} & \colhead{[Ca/Fe]} &
\colhead{[Sc/Fe]} & \colhead{[Ti/Fe]}} \startdata
BD+04\degr~2466  & $-1.92\pm0.05$ & $ 0.16\pm0.08$ & $ 0.25\pm0.08$ & $ 0.50\pm0.07$ & $ 0.45\pm0.08$ & $ 0.17\pm0.08$ & $0.37\pm0.08$ \\
BD+01\degr~3070  & $-1.52\pm0.05$ & $-0.09\pm0.06$ & $ 0.32\pm0.09$ & $ 0.36\pm0.09$ & $ 0.24\pm0.06$ & $ 0.23\pm0.05$ & $0.23\pm0.06$ \\
BD+09\degr~2870  & $-2.73\pm0.05$ & $ 0.11\pm0.09$ & $ 0.46\pm0.13$ & $    ...     $ & $ 0.46\pm0.10$ & $ 0.04\pm0.05$ & $0.23\pm0.11$ \\
BD+10\degr~2495  & $-2.10\pm0.05$ & $ 0.06\pm0.06$ & $ 0.22\pm0.08$ & $ 0.45\pm0.05$ & $ 0.34\pm0.06$ & $ 0.12\pm0.06$ & $0.17\pm0.06$ \\
BD+12\degr~2547  & $-1.89\pm0.05$ & $-0.15\pm0.06$ & $ 0.36\pm0.08$ & $ 0.30\pm0.09$ & $ 0.23\pm0.06$ & $ 0.2 \pm0.07$ & $0.10\pm0.06$ \\
BD+29\degr~2356  & $-1.45\pm0.05$ & $-0.04\pm0.09$ & $ 0.36\pm0.08$ & $ 0.35\pm0.07$ & $ 0.25\pm0.07$ & $ 0.25\pm0.07$ & $0.24\pm0.06$ \\
BD+30\degr~2611  & $-1.41\pm0.06$ & $-0.33\pm0.08$ & $ 0.14\pm0.08$ & $ 0.24\pm0.08$ & $ 0.05\pm0.07$ & $    ...     $ & $0.05\pm0.07$ \\
HD~33771         & $-2.06\pm0.05$ & $-0.04\pm0.06$ & $ 0.29\pm0.08$ & $ 0.42\pm0.07$ & $ 0.30\pm0.06$ & $ 0.16\pm0.08$ & $0.19\pm0.05$ \\
HD~85773         & $-2.58\pm0.05$ & $-0.71\pm0.08$ & $ 0.29\pm0.09$ & $    ...     $ & $ 0.24\pm0.07$ & $-0.02\pm0.06$ & $0.28\pm0.07$ \\
HD~107752        & $-3.16\pm0.05$ & $-0.47\pm0.05$ & $ 0.38\pm0.10$ & $    ...     $ & $ 0.54\pm0.06$ & $-0.14\pm0.08$ & $0.13\pm0.07$ \\
HD~108577        & $-2.37\pm0.05$ & $ 0.05\pm0.05$ & $ 0.42\pm0.08$ & $    ...     $ & $ 0.40\pm0.06$ & $ 0.13\pm0.07$ & $0.27\pm0.05$ \\
HD~119516        & $-1.82\pm0.05$ & $ 0.16\pm0.07$ & $ 0.16\pm0.09$ & $    ...     $ & $ 0.22\pm0.07$ & $ 0.08\pm0.07$ & $0.31\pm0.10$ \\
HD~124358        & $-1.77\pm0.05$ & $-0.28\pm0.06$ & $    ...     $ & $ 0.34\pm0.05$ & $ 0.15\pm0.07$ & $ 0.04\pm0.07$ & $0.04\pm0.06$ \\
HD~128279        & $-2.24\pm0.05$ & $-0.38\pm0.09$ & $ 0.44\pm0.07$ & $ 0.47\pm0.08$ & $ 0.35\pm0.07$ & $ 0.22\pm0.07$ & $0.30\pm0.05$ \\
HD~175305        & $-1.35\pm0.09$ & $-0.02\pm0.07$ & $ 0.34\pm0.08$ & $ 0.31\pm0.10$ & $ 0.23\pm0.07$ & $ 0.35\pm0.10$ & $0.24\pm0.07$ \\
HD~237846        & $-3.01\pm0.05$ & $-0.39\pm0.05$ & $ 0.47\pm0.10$ & $    ...     $ & $ 0.37\pm0.08$ & $-0.04\pm0.09$ & $0.27\pm0.06$ \\
HD~134439        & $-1.35\pm0.05$ & $-0.44\pm0.06$ & $ 0.08\pm0.12$ & $ 0.05\pm0.05$ & $ 0.01\pm0.05$ & $ 0.08\pm0.09$ & $0.25\pm0.06$ \\
G~112--43        & $-1.31\pm0.05$ & $-0.03\pm0.06$ & $ 0.35\pm0.12$ & $ 0.26\pm0.06$ & $ 0.28\pm0.06$ & $ 0.24\pm0.05$ & $0.40\pm0.05$ \\
G~115--58        & $-1.36\pm0.05$ & $-0.43\pm0.05$ & $ 0.15\pm0.08$ & $ 0.11\pm0.07$ & $ 0.23\pm0.07$ & $ 0.05\pm0.07$ & $0.30\pm0.06$ \\
G~15--13         & $-1.63\pm0.05$ & $-0.37\pm0.06$ & $ 0.21\pm0.08$ & $    ...     $ & $ 0.09\pm0.07$ & $ 0.15\pm0.08$ & $0.24\pm0.06$ \\
G~166--37        & $-1.36\pm0.08$ & $-0.22\pm0.05$ & $ 0.25\pm0.07$ & $ 0.18\pm0.07$ & $ 0.15\pm0.05$ & $ 0.34\pm0.06$ & $0.17\pm0.06$ \\
G~238--30        & $-3.44\pm0.06$ & $-0.65\pm0.06$ & $ 0.74\pm0.08$ & $    ...     $ & $ 0.69\pm0.06$ & $ 0.04\pm0.20$ & $0.49\pm0.10$ \\
G~41--41         & $-2.84\pm0.06$ & $-0.71\pm0.07$ & $ 0.36\pm0.08$ & $    ...     $ & $ 0.39\pm0.06$ & $ 0.34\pm0.10$ & $0.59\pm0.09$ \\
G~48--29         & $-2.82\pm0.05$ & $-0.50\pm0.06$ & $ 0.56\pm0.08$ & $    ...     $ & $ 0.58\pm0.09$ & $ 0.26\pm0.08$ & $0.56\pm0.10$ \\
G~53--41         & $-1.22\pm0.05$ & $ 0.25\pm0.08$ & $ 0.22\pm0.08$ & $ 0.23\pm0.07$ & $ 0.20\pm0.06$ & $ 0.18\pm0.07$ & $0.26\pm0.08$ \\
LP~894--3        & $-1.50\pm0.05$ & $-0.17\pm0.06$ & $ 0.14\pm0.08$ & $ 0.28\pm0.05$ & $ 0.14\pm0.06$ & $ 0.05\pm0.09$ & $0.18\pm0.11$ \\

\enddata
\end{deluxetable}

\clearpage
\begin{deluxetable}{lcrrrrrr}
\tabletypesize{\scriptsize} \rotate \tablecaption{Abundance relative
to Fe: Cr, Mn, Ni, Zn, Y, and Ba} \tablewidth{0pt}
\tablehead{\colhead{Star} & \colhead{[Fe/H ]} & \colhead{[Cr/Fe]} &
\colhead{[Mn/Fe]} & \colhead{[Ni/Fe]} & \colhead{[Zn/Fe]} &
\colhead{[Y/Fe]} & \colhead{[Ba/Fe]}} \startdata
BD+04\degr~2466   & $-1.92\pm0.05$ & $-0.10\pm0.08$ & $-0.44\pm0.04$ & $ 0.02\pm0.08$ & $-0.04\pm0.10$ & $ 0.55\pm0.05$ & $ 1.31\pm0.05$ \\
BD+01\degr~3070   & $-1.52\pm0.05$ & $-0.14\pm0.06$ & $-0.58\pm0.03$ & $-0.11\pm0.07$ & $-0.06\pm0.03$ & $-0.14\pm0.05$ & $-0.07\pm0.06$ \\
BD+09\degr~2870   & $-2.73\pm0.05$ & $-0.32\pm0.05$ & $-0.61\pm0.07$ & $-0.01\pm0.07$ & $-0.05\pm0.03$ & $-0.48\pm0.05$ & $-0.80\pm0.06$ \\
BD+10\degr~2495   & $-2.10\pm0.05$ & $-0.18\pm0.09$ & $-0.49\pm0.03$ & $-0.07\pm0.07$ & $-0.01\pm0.03$ & $-0.26\pm0.05$ & $-0.11\pm0.06$ \\
BD+12\degr~2547   & $-1.89\pm0.05$ & $-0.20\pm0.07$ & $-0.60\pm0.03$ & $-0.06\pm0.07$ & $-0.07\pm0.03$ & $-0.14\pm0.05$ & $-0.10\pm0.06$ \\
BD+29\degr~2356   & $-1.45\pm0.05$ & $-0.12\pm0.07$ & $-0.67\pm0.08$ & $-0.09\pm0.08$ & $-0.08\pm0.03$ & $-0.05\pm0.05$ & $ 0.01\pm0.06$ \\
BD+30\degr~2611   & $-1.41\pm0.06$ & $-0.02\pm0.10$ & $-0.65\pm0.04$ & $-0.15\pm0.09$ & $-0.32\pm0.03$ & $-0.22\pm0.05$ & $-0.05\pm0.06$ \\
HD~33771          & $-2.06\pm0.05$ & $-0.24\pm0.05$ & $-0.54\pm0.04$ & $-0.07\pm0.07$ & $-0.08\pm0.04$ & $-0.15\pm0.05$ & $-0.09\pm0.06$ \\
HD~85773          & $-2.58\pm0.05$ & $-0.27\pm0.07$ & $-0.44\pm0.06$ & $ 0.00\pm0.08$ & $ 0.44\pm0.05$ & $-0.57\pm0.05$ & $-0.69\pm0.06$ \\
HD~107752         & $-3.16\pm0.05$ & $-0.23\pm0.07$ & $-0.55\pm0.03$ & $-0.02\pm0.08$ & $ 0.05\pm0.04$ & $-0.40\pm0.06$ & $-0.56\pm0.06$ \\
HD~108577         & $-2.37\pm0.05$ & $-0.15\pm0.07$ & $-0.31\pm0.04$ & $ 0.08\pm0.07$ & $ 0.10\pm0.04$ & $-0.32\pm0.08$ & $-0.17\pm0.06$ \\
HD~119516         & $-1.82\pm0.05$ & $-0.15\pm0.08$ & $-0.41\pm0.04$ & $-0.07\pm0.10$ & $-0.06\pm0.07$ & $-0.14\pm0.05$ & $ 0.11\pm0.06$ \\
HD~124358         & $-1.77\pm0.05$ & $-0.24\pm0.06$ & $-0.44\pm0.03$ & $-0.17\pm0.07$ & $-0.28\pm0.04$ & $-0.60\pm0.05$ & $-0.43\pm0.06$ \\
HD~128279         & $-2.24\pm0.05$ & $-0.18\pm0.05$ & $-0.42\pm0.03$ & $ 0.04\pm0.08$ & $-0.03\pm0.06$ & $    ...     $ & $-0.73\pm0.07$ \\
HD~175305         & $-1.35\pm0.09$ & $-0.12\pm0.07$ & $-0.61\pm0.05$ & $-0.09\pm0.07$ & $-0.03\pm0.04$ & $ 0.00\pm0.09$ & $-0.01\pm0.06$ \\
HD~237846         & $-3.01\pm0.05$ & $-0.24\pm0.06$ & $-0.48\pm0.03$ & $-0.04\pm0.12$ & $ 0.06\pm0.04$ & $-0.51\pm0.07$ & $-1.08\pm0.10$ \\
HD~134439         & $-1.35\pm0.05$ & $-0.09\pm0.07$ & $-0.51\pm0.03$ & $-0.12\pm0.08$ & $-0.10\pm0.03$ & $-0.30\pm0.05$ & $-0.24\pm0.06$ \\
G~112--43         & $-1.31\pm0.05$ & $-0.06\pm0.05$ & $-0.27\pm0.03$ & $ 0.02\pm0.05$ & $ 0.19\pm0.04$ & $-0.09\pm0.05$ & $-0.30\pm0.06$ \\
G~115--58         & $-1.36\pm0.05$ & $-0.11\pm0.05$ & $-0.42\pm0.03$ & $-0.13\pm0.10$ & $-0.21\pm0.07$ & $-0.08\pm0.05$ & $-0.15\pm0.10$ \\
G~15--13          & $-1.63\pm0.05$ & $-0.06\pm0.06$ & $-0.40\pm0.10$ & $-0.08\pm0.08$ & $-0.02\pm0.04$ & $-0.03\pm0.06$ & $ 0.22\pm0.09$ \\
G~166--37         & $-1.36\pm0.08$ & $-0.06\pm0.06$ & $-0.48\pm0.07$ & $-0.07\pm0.03$ & $-0.26\pm0.05$ & $ 0.09\pm0.06$ & $ 0.16\pm0.10$ \\
G~238--30         & $-3.44\pm0.06$ & $-0.13\pm0.09$ & $    ...     $ & $    ...     $ & $    ...     $ & $     ...    $ & $    ...     $ \\
G~41--41          & $-2.84\pm0.06$ & $    ...     $ & $    ...     $ & $    ...     $ & $    ...     $ & $     ...    $ & $-0.36\pm0.06$ \\
G~48--29          & $-2.82\pm0.05$ & $-0.07\pm0.10$ & $    ...     $ & $    ...     $ & $    ...     $ & $     ...    $ & $-0.54\pm0.08$ \\
G~53--41          & $-1.22\pm0.05$ & $-0.13\pm0.06$ & $-0.41\pm0.03$ & $-0.07\pm0.07$ & $-0.06\pm0.05$ & $ 0.10\pm0.05$ & $ 0.27\pm0.06$ \\
LP~894--3         & $-1.50\pm0.05$ & $-0.15\pm0.05$ & $-0.45\pm0.03$ & $-0.09\pm0.10$ & $-0.24\pm0.03$ & $     ...    $ & $-0.20\pm0.06$ \\
\enddata
\end{deluxetable}


\begin{thebibliography}{}

\bibitem[Alonso et al.(1996)]{alo96} Alonso, A., Arribas, S.,
Mart\'inez-Roger, C., 1996, \aap, 313, 873
\bibitem[Alonso et al.(1999)]{alo99} Alonso, A., Arribas, S.,
Mart\'inez-Roger, C., 1999, \aaps, 140, 261
\bibitem[Anders \& Grevesse(1989)]{and89} Anders, E.,
\& Grevesse, N., 1989, \gca, 53, 197
\bibitem[Andrievsky et al.(2007)]{and07} Andrievsky, S. M., Spite, M., Korotin, S. A., \& Spite, F.
et al., \aap, 464, 1081
\bibitem[Aoki et al.(2005)]{aok05} Aoki, W., Honda, S., Beers, T. C.,
et al., 2005, \apj, 632, 611
\bibitem[Aoki et al.(2009a)]{aok09} Aoki, W., Barklem, P.~S.,
Beers, T.~C., Christlieb, N., Inoue, S., Garcia Perez, A.~E.,
Norris, J.~E., \& Carollo, D., 2009, \apj, 698, 1803
\bibitem[Aoki et al.(2009b)]{aok09b} Aoki, W., et al., 2009, \aap,
502, 569
\bibitem[Barklem et al.(2005)]{bar05} Barklem, P. S., Christlieb, N., Beers, T. C., et al., 2005,
\aap, 439, 129
\bibitem[Baum\"{u}ller et al.(1998)]{bau98} Baum\"{u}ller, D.,
Butler, K., \& Gehren, T., 1998, \aap, 338, 637
\bibitem[Beers et al.(2000)]{bee00} Beers, T. C., Chiba, M.,
Yoshii, Y., Platais, I., Hanson, R. B., Fuchs, B. \& Rossi, S.,
2000, AJ, 119, 2866
\bibitem[Beers \& Christlieb(2005)]{bee05} Beers, T. C., \& Christlieb,
N., 2005, \araa, 43, 531
\bibitem[Booth et al.(1984)]{boo84} Booth, A. J., Blackwell, D. E., Petford, A. D., \& Shallis, M. J., 1984,
\mnras, 208, 147
\bibitem[Bi\'{e}mont \& Godefroid(1980)]{bie80} Bi\'{e}mont, E., \& Godefroid, M. 1980, \aap, 84, 361
\bibitem[Bonifacio et al.(2009)] {bon09} Bonifacio, P., Spite, M., \& Cayrel, R.et al.,  2009, arXiv:
0903.4174
\bibitem[Bullock \& Johnston(2005)]{bul05} Bullock, J. S., \&
Johnston, K. V., 2005, \apj, 635, 931
\bibitem[Busso et al.(2001)]{bus01} Busso, M., Gallino, R., Lambert, D.L.,
Travaglio, C., \&Smith, V.V., 2001, \apj, 557, 802
\bibitem[Carney et al.(1994)]{car94} Carney, B. W., Latham, D. W., Laird, J. B., \& Aguilar, L.
A., 1994, \aj, 107, 2240
\bibitem[Carney et al. (2003)]{car03} Carney, B. W., Latham, D. W., Stefanik, R. P., Laird, J. B., \& Morse, J.
A., 2003, \aj, 125, 293
\bibitem[Carretta et al.(2002)]{car02} Carretta, E., Gratton, R.,
Cohen, J., et al., 2002, \apj, 124, 481
\bibitem[Castelli et al.(1997)]{cas97} Castelli, F., Gratton, R. G., \& Kurucz, R. L., 1997, \aap, 318, 841
\bibitem[Castelli \& Kurucz(2003)]{castelli03} Castelli, F., \&
Kurucz, R.~L.\ 2003, Modelling of Stellar Atmospheres, 210, 20P
\bibitem[Carollo et al. (2007)]{car07} Carollo, D., et al., 2007, \nat, 450, 1020
\bibitem[Cayrel(1988)]{cay88} Cayrel, R. 1988, in The impact of very high S/N spectroscopy on
Stellar Physics, ed. G. Cayrel de Strobel, \& M. Spite (Kluwer),
Proc. IAU Symp., 132, 345
\bibitem[Cayrel et al.(2004)]{cay04} Cayrel, R., Depagne, E., Spite,
M., et al. 2004, \aap, 416, 1117
\bibitem[Chiba \& Beers(2000)]{chi00} Chiba, M., \& Beers, T. C.,
2000, \aj, 549, 325
\bibitem[Cohen et al.(2002)]{coh02} Cohen, J. G., Christlieb, N., Beers, T.
C., et al., 2002, \apj, 124, 470
\bibitem[Frebel et al.(2009)]{fre09} Frebel, A., Simon, J. D., Geha, M., \& Willman, B., 2009, arXiv:0902.2395
\bibitem[Fulbright(2000)]{ful00} Fulbright, J. P., 2000, \aj, 120,
1841
\bibitem[Fulbright(2002)]{ful02} Fulbright, J. P., 2002, \aj, 123,
404
\bibitem[Honda et al.(2004)]{hon04} Honda, S., Aoki, W., Kajino, T.,
et al., 2004, \apj, 607, 474
\bibitem[Heger \& Woosley (2002)] {heg02} Heger, A., \& Woosley, S.
E., 2002, \apj, 567, 532
\bibitem[Gehren et al.(2004)]{geh04} Gehren, T., Liang, Y. C., Shi,
J. R., et al., 2004, \aap, 413, 1045
\bibitem[Gehren et al.(2006)]{geh06} Gehren, T., Shi, J. R., Zhang,
H. W., et al., 2006, \aap, 451, 1065
\bibitem[Gilmore \& Wyse(1998)]{gil98} Gilmore, G., \& Wyse, R. F. G., 1998, \aj,
116, 748
\bibitem[Gratton et al.(1997)]{gra97} Gratton, R. G., Carretta, E., Clementini, G., \& Sneden, C. 1997, in Proc. of
the ESA Symposium: Hipparcos¡ªVenice '97, ESA SP-402 (Noordwijk:
ESA), 339
\bibitem[Gratton et al.(2003)]{gra03} Gratton, R. G., Carretta, E., Claudi, R., Lucatello, S., \& Barbieri,
M, 2003, \aap, 404, 187
\bibitem[Ibata et al.(1994)]{iba94} Ibata, R. A., Gilmore, G., \& Irwin, M. J., 1994, \nat, 370, 194
\bibitem[Jorissen et al.(2005 )]{jor05} Jorissen, A., Zacs, L., Udry, S., Lindgren, H., \& Musaev, F.
A., 2005, \aap, 441, 1135
\bibitem[Johnston et al.(2008)]{joh08} Johnston, K. V., Bullock, J. S., Sharma, S., Font, A., Robertson, B. E., \& Leitner, S.
N., 2008, \apj, 689, 936
\bibitem[Kobayashi et al.(2006)]{kob06} Kobayashi, C., Umeda, H.,
Nomoto, K., Tominaga, N., \& Ohkubo, T., 2006, \apj, 653, 1145
\bibitem[Kurucz(1993)]{kur93} Kurucz, R. L., 1993, CD-ROM No. 13, 18, Smithsonian Atrophysical
Observatory
\bibitem[Lai et al.(2008)]{lai08} Lai, D. K., Bolte, M., Johnson, Jennifer A., \& Lucatello, S.,
et al., 2008, \apj, 681, 1524
\bibitem[Lanfranchi \& Matteucci(2004)]{lan04} Lanfranchi, G. A., \&
Matteucci, F., 2004, \mnras, 351, 1338
\bibitem[Latham et al.(2002)]{lat02} Latham, D. W., Stefanik, R. P., \& Torres,
G. et al., 2002 \apj, 124, 1144
\bibitem[Lawler \& Dakin(1989)]{law89} Lawler, J. E., \&
Dakin, J. T., 1989, J. Opt. Soc. Am. B 6, 1457
\bibitem[McWilliam et al.(1995)]{mcw95} McWilliam, A., Preston, G. W., Sneden, C., \& Searle,
L., 1995, \aj, 109, 2757
\bibitem[McWilliam(1998)]{mcw98} McWilliam A., 1998, \aj, 115, 1640
\bibitem[Noguchi et al.(2002)]{nog02} Noguchi, K., Aoki, W., Kawanomoto, S., Ando,
H., et al., 2002, \pasj, 54, 855
\bibitem[Norris(1994)]{nor94} Norris, J. E. 1994, ApJ, 431, 645
\bibitem[Perryman et al.(1997)]{per97} Perryman, M. A. C., Lindegren, L., Kovalevsky, J., \& Hoeg, E., et al.,
1997, \aap, 323L, 49
\bibitem[Preston et al.(1994)]{pre94} Preston, G. W., Beers, T. C., \& Schectman, S. A., 1994, \aj, 108, 538
\bibitem[Ramirez \& Mel\'{e}nclez(2004)]{ram04} Ramirez, I., \&
Mel\'{e}nclez, J., 2004, \apj, 609, 417
\bibitem[Roederer(2009)]{roe09} Reoderer, I. U., 2009, \apj, 137,
272
\bibitem[Ryan \& Norris(1991)]{rya91} Ryan, S. G., \& Norris, J. E.,
1991, \aj, 101, 1865
\bibitem[Ryan et al.(1996)]{rya96} Ryan, S. G., Norris, J. E., Beers, T. C., 1996, \apj, 471, 254
\bibitem[Schlegel et al.(1998)]{sch98} Schlegel, D. J., Finkbeiner, D. P., \& Davis, M., 1998, \apj, 500, 525
\bibitem[Shetron et al.(2001)]{she01} Shetron, M., C\^{o}t\'{e}, P.,
\& Sargent, W. L. W., 2001, \apj, 548, 592
\bibitem[Shetron et al.(2003)]{she03} Shetron, M., Venn, K. M.,
Tolstoy, E., Primas, F., Hill, V., \& Kaufer, A., 2003, \apj, 125,
684
\bibitem[Shigeyama \& Tsujimoto(1998)]{shi98} Shigeyama,T., \&
Tsujimoto, T., 1998, \apj, 507, L135
\bibitem[Sneden et al.(1996)]{sne96} Sneden, C., McWilliam, A., Preston, G. W.,
et al., 1996, \apj, 467, 819
\bibitem[Sobeck et al.(2007)]{sob07} Sobeck. J. S., Lawler, J. E.,
\& Sneden, C., 2007, \apj, 667, 1267
\bibitem[Sommer-Larsen \& Zhen(1990)]{som90} Sommer-Larsen, J. \& Zhen, C., 1990, MNRAS, 242, 10
\bibitem[Stephens(1999)]{ste99} Stephens, A., 1999, \aj, 117, 1771
\bibitem[Stephens \& Boesgaard(2002)]{ste02} Stephens, A., \&
Boesgaard, A. M., 2002, \aj, 123, 1647
\bibitem[Tsujimoto et al.(1995)]{tsu95} Tsujimoto, T., Nomoto, K., Yoshii, Y., Hashimoto, M., Yanagida, S., \&
Thielemann, F.-K., 1995, \mnras, 277, 94
\bibitem[Umeda \& Nomoto(2005)]{ume05} Umeda, H., \& Nomoto, K.,
2005, \apj, 619, 427
\bibitem[Woosley \& Weaver(1995)]{woo95} Woosley, S.E., \& Weaver,
T. A., 1995, \apjs, 101, 181
\bibitem[Zhao \& Magain(1990)]{zhao90} Zhao, G., \& Magain, P., 1990,
\aap, 238, 242
\bibitem[Zhao \& Magain(1991)]{zhao91} Zhao, G., \& Magain, P.,
1991, \aap, 244, 425

\end{thebibliography}
\end{document}